\documentclass[amsfonts,showpacs,tightenlines,aps,12pt,floatfix]{revtex4}
\usepackage{bm}
\usepackage{epsfig}
\begin{document}


\title{$\alpha_s(M_Z^2)$ From Hadronic $\tau$ Decays}
\author{K. Maltman}
\email[]{kmaltman@yorku.ca}
\affiliation{Department of Mathematics and Statistics, York University, 
4700 Keele St., Toronto, ON CANADA M3J 1P3}
\altaffiliation{CSSM, Univ. of Adelaide, Adelaide, SA 5005 AUSTRALIA}
\author{T. Yavin}
\email[]{t_yavin@yorku.ca}
\affiliation{Department of Physics and Astronomy, York University, 
4700 Keele St., Toronto, ON CANADA M3J 1P3}

\date{\today}

\begin{abstract}
We perform an extraction of $\alpha_s$ based on sum rules 
involving isovector hadronic $\tau$ decay data. The particular
sum rules employed are constructed specifically to suppress 
contributions associated with poorly known higher dimension condensates, 
and hence reduce theoretical systematic uncertainties associated with the 
treatment of such contributions which are shown to be present
in earlier related analyses. Running
our results from the $n_f=3$ to $n_f=5$ regime we find
$\alpha_s(M_Z^2)=0.1187\pm 0.0016$, in excellent agreement with
the recently updated global fit to electroweak data at the Z scale 
and other high-scale direct determinations.
\end{abstract}

\pacs{12.38.-t,13.35.Dx,11.55.Hx}

\maketitle

\section{\label{intro}Introduction}
The value of the running strong coupling, $\alpha_s(\mu^2 )$,
at some conventionally chosen reference scale
is one of the fundamental parameters of the Standard Model (SM). 
In what follows, we adhere to standard
convention and quote results at the scale $\mu =M_Z$, for $n_f=5$,
in the $\overline{MS}$ scheme, and denote this quantity
by $\alpha_s(M_Z^2)$. 

The running coupling $\alpha_s(\mu^2 )$ has been determined
experimentally in a large number of independent processes, over a wide range
of scales~\cite{pdg06qcdreview}. The observed variation, by a factor of 
$\sim 3$, over the range from $\mu \sim 2$ GeV to $\mu =M_Z$ is 
in excellent agreement with QCD expectations, and represents a
highly non-trivial test of the theory. If, however, one looks
in more detail, one finds that the two highest-precision low-energy 
determinations, that coming from a lattice perturbation
theory analysis of UV-sensitive lattice observables~\cite{latticealphas}, 
and that coming from finite energy sum rule (FESR) analyses of 
hadronic $\tau$ decay data~\cite{dhzreview,bck08,davieretal08}, 
are not in good agreement within their mutual errors, 
the most recent determinations yielding
\begin{eqnarray}
\alpha_s(M_Z^2)\,&& =\, 0.1170\pm 0.0012\ {\rm (lattice)}
\label{hpqcdukqcdalphas}\\
\alpha_s(M_Z^2)\,&& =\, 0.1212\pm 0.0011\ {\rm (}\tau {\rm\  decay)}
\label{taualphas}\end{eqnarray}
for the lattice~\cite{latticealphas} and $\tau$ decay~\cite{davieretal08}
determinations, respectively.

In this paper we revisit the hadronic $\tau$ decay extraction,
focussing on alternate FESR choices designed specifically
to reduce theoretical systematic
uncertainties {\it not included in the error assessment of
Eq.~(\ref{taualphas})} and associated with possible small higher
dimension ($D>8$) OPE contributions assumed negligible in the analyses
reported in Refs.~\cite{dhzreview,davieretal08}. We find a shift in the
results for $\alpha_s(M_Z^2)$ in excess of the previously quoted error,
and obtain also an improvement in the
agreement (i) between the $\tau$ decay and direct
high-scale determinations and
(ii) amongst the separate $\tau$ decay extractions obtained from
the vector (V), axial vector (A), and vector-plus-axial-vector 
(V+A) channel analyses. 

The rest of the paper is organized as follows. In Section~\ref{sec2}
we (i) outline the general FESR approach to extracting $\alpha_s$
from hadronic $\tau$ decay data, (ii) discuss the relevant features 
of existing analyses, (iii) point out potential additional theoretical 
uncertainties in those analyses, associated with the neglect of
$D>8$ OPE contributions, (iv) establish explicitly the presence
of such contributions at a level {\it not} negligible
on the scale of the previously quoted errors, and (v) discuss 
alternate sum rule choices which significantly 
reduce these uncertainties. In Section~\ref{sec3} we use these alternate
sum rules to perform separate V, A and V+A analyses, employing either
the ALEPH~\cite{dhzreview,davieretal08,alephud05,alephud9798} 
or OPAL~\cite{opalud99} isovector hadronic $\tau$ decay data sets. 
Our final results for $\alpha_S(M_Z^2)$, together with a discussion
of these results, are given in Section~\ref{sec4}.

\section{\label{sec2}Hadronic $\tau$ decay extractions of $\alpha_s$}

\subsection{The Finite Energy Sum Rule Framework}
The kinematics of $\tau$ decay in the SM allows the inclusive rate
for hadronic $\tau$ decays mediated by the flavor $ij=ud,us$, 
V or A currents to be written as a sum of kinematically weighted integrals
over the spectral functions $\rho_{V/A;ij}^{(J)}(s)$,
associated with the spin $J=0,1$ components of the relevant
current-current two-point functions~\cite{tsaitaubasic}. 
Defining $R_{V/A;ij}\equiv  \Gamma [\tau^- \rightarrow \nu_\tau
\, {\rm hadrons}_{V/A;ij}\, (\gamma)]/
\Gamma [\tau^- \rightarrow
\nu_\tau e^- {\bar \nu}_e (\gamma)]$
and $y_\tau\, \equiv\, s/m_\tau^2$, 
one has
\begin{equation}
R_{V/A;ij}= 12\pi^2\vert V_{ij}\vert^2 S_{EW}\,
\int^{1}_0\, dy_\tau \,\left( 1-y_\tau\right)^2
\left[ \left( 1 + 2y_\tau\right)
\rho_{V/A;ij}^{(0+1)}(s) - 2y_\tau \rho_{V/A;ij}^{(0)}(s) \right]
\label{taukinspectral}
\end{equation}
with $V_{ij}$ the flavor $ij$ CKM matrix element, $S_{EW}$ a short-distance 
electroweak correction~\cite{bl90,erler,sewfootnote}, 
and $\rho_{V/A;ij}^{(0+1)}(s)
\equiv\rho_{V/A;ij}^{(1)}(s)+\rho_{V/A;ij}^{(0)}(s)$.
We concentrate here on the isovector ($ij=ud$) case.

For $ij=ud$, apart from the $\pi$ pole contribution to
$\rho_{A;ud}^{(0)}$, all contributions to $\rho_{V;ud}^{(0)}(s)$, 
$\rho_{A;ud}^{(0)}(s)$, are of $O([m_d\mp m_u]^2)$,
and hence numerically negligible, allowing the sum of
the flavor $ud$ V and A spectral functions $\rho_{V+A;ud}^{(0+1)}(s)$ 
to be determined directly from experimental results for $dR_{V+A;ud}/ds$. 
Further separation into V and A components is unambiguous for
$n\pi$ states, but requires additional input for $K\bar{K}n\pi$ ($n>0$)
states. Errors on the experimental
distribution are thus reduced by working with the V+A sum.

The spectral functions, $\rho_{V/A;ij}^{(0+1)}(s)$, correspond to
scalar correlator combinations, 
$\Pi_{V/A;ij}^{(0+1)}(s)\equiv \Pi_{V/A;ij}^{(1)}(s)+\Pi_{V/A;ij}^{(0)}(s)$,
having no kinematic singularities. For any such correlator, 
$\Pi (s)$, with spectral function $\rho (s)$, and any $w(s)$ analytic 
in $\vert s\vert <M$ with $M>s_0$, analyticity implies the finite
energy sum rule (FESR) relation
\begin{equation}
\int_0^{s_0}w(s)\, \rho(s)\, ds\, =\, -{\frac{1}{2\pi i}}\oint_{\vert
s\vert =s_0}w(s)\, \Pi (s)\, ds\ .
\label{basicfesr}
\end{equation}
For sufficiently large $s_0$, the OPE representation
can be employed on the RHS of Eq.~(\ref{basicfesr}). The region of 
applicability of the OPE is extended to lower $s_0$ by working with 
``pinched'' weights (those satisfying $w(s=s_0)=0$), which suppress 
contributions on the RHS from the region of the contour near the timelike real
axis~\cite{pqw,kmfesr}. 

For FESRs employed hadronic $\tau$ decay data, $s_0$ up to $m_\tau^2$
are kinematically allowed on the RHS of Eq.~(\ref{basicfesr}).
Since $m_\tau = 1.77684(17)$ GeV~\cite{pdg08} is
$>>\Lambda_{QCD}$, one expects
the integrated OPE to provide a reliable representation over 
a significant portion of the kinematically allowed $s_0$ range.

In previous extractions of $\alpha_s$, FESRs 
involving $\Pi_T (s)\equiv \Pi^{(0+1)}_{T;ud}(s)$ (with T=V, A or V+A), 
pinched polynomial weights, and $s_0=m_\tau^2$ were employed. 
Our analysis will employ a range of $s_0$ and an alternate set of such 
weights having the generic form $w(y)=\sum_m b_my^m$, with 
$y=s/s_0$~\cite{dimlessfootnote}. 

\subsection{Experimental Input for the Weighted Spectral Integrals}
Data and covariance matrices for the spectral distributions
$dR_{T;ud}/ds$, again with $T=V,A$ and $V+A$, have been
provided by both the ALEPH~\cite{dhzreview,alephud05,alephud9798} 
and OPAL~\cite{opalud99} collaborations. The ALEPH covariances
lead to weighted spectral integrals with 
non-normalization-induced errors a factor of $\sim 2$ smaller
than those obtained using the OPAL results. 

In addition, 
ALEPH has recently provided previously unavailable information
on the V+A $K\bar{K}\pi$ distribution~\cite{davieretal08}, a mode 
for which separate information is not available from OPAL. This is 
of relevance to performing the separate V and A analyses since recent BaBar 
determinations of the isovector 
$K\bar{K}\pi$ electroproduction cross-sections~\cite{babarkkbarpi07},
combined with CVC, allow for a significant improvement in the treatment of 
the V/A separation in the $K\bar{K}\pi$ channel~\cite{davieretal08},
which channel dominates the uncertainty in the V/A separation for
non-strange hadronic $\tau$ decays. In view of these advantages, 
we will focus our discussion on the ALEPH data~\cite{alephkkbarpifootnote}, 
though we will also perform alternate independent analyses using 
the OPAL data as input, as a further consistency check. 

\subsection{The OPE Representation of $\Pi_{V/A;ud}$}

\subsubsection{The $D=0$ Contribution}
On the OPE side of Eq.~(\ref{basicfesr}), for most weights $w(s)$,
and for scales above $s_0\sim 2\ {\rm GeV}^2$, 
far and away the dominant contribution comes 
from the $D=0$ term, which is conveniently written in terms 
of the Adler function, $D_T(Q^2)\equiv -Q^2\, d\Pi_T (Q^2)/dQ^2$, 
\begin{equation}
\oint_{\vert s\vert =s_0}ds\, w(s)\, \left[\Pi_T (s)\right]_{D=0}\, =\, 
\oint_{\vert s\vert =s_0}ds\, {\frac{v(s)}{s}}\, 
\left[ D_T (Q^2)\right]_{D=0}\, , 
\label{d0ope}
\end{equation}
where $Q^2\, =\, -s$ and $v(s)=\int\, ds\, w(s)$, with $v(s_0)=0$. 
In this form, potentially large logarithms can be summed up point-by-point 
along the contour through the scale choice $\mu^2\, =\, Q^2$.
The resulting ``contour-improved'' (CIPT) evaluation 
improves the convergence behavior of the 
known terms of the integrated $D=0$ series~\cite{cipt}. 
An alternate evaluation, referred to as ``fixed order perturbation
theory'' (FOPT), involves choosing a common fixed scale 
(such as $\mu^2 =s_0$)
for all points on the contour. Large logarithms are then unavoidable
over at least some portion of the contour. Detailed arguments 
in favor of the CIPT prescription have been presented in
Ref.~\cite{davieretal08}. We find optimal consistency
of our results when employing the CIPT implementation,
and thus take the CIPT evaluation as our central one. 
However, the difference between the CIPT and FOPT
evaluations, both truncated at the same given order, 
lies entirely in contributions of yet higher order. 
The CIPT-FOPT difference thus serves as one possible measure of the $D=0$ 
series truncation uncertainty. It turns out that this difference is, 
in most cases, significantly larger than other possible estimates of the
same uncertainty. We will thus adopt a conservative view and
include the full CIPT-FOPT difference as one component of our 
truncation uncertainty estimate.

The $D=0$ contribution to $D_{V/A;ij}$ is known to $O(\alpha_s^4)$, 
and given by
\begin{equation}
\left[ D_{V/A;ij}(Q^2)\right]_{D=0}\, =\, {\frac{1}{4\pi^2}}\, 
\sum_{k\geq 0}d^{(0)}_k \bar{a}^k\ ,
\label{dzeroadler}\end{equation}
where $\bar{a}=a(Q^2)=\alpha_s(Q^2)/\pi$, with
$\alpha_s(\mu^2)$ the running coupling at scale $\mu^2$
in the $\overline{MS}$ scheme, and, for $n_f=3$, 
$d^{(0)}_0=d^{(0)}_1=1$, 
$d^{(0)}_2=1.63982$, $d^{(0)}_3=6.37101$ and
$d^{(0)}_4=49.07570$~\cite{d0adler,bck08}.
The next coefficient, $d^{(0)}_5$, has been estimated to be 
$\sim 275$~\cite{bck08} using methods known to have (i) worked well 
semi-quantitatively for the coefficients of the $D=0$ series~\cite{bck02} and
(ii) produced, in advance of the actual calculation, 
an accurate prediction for the recently computed $O(a^3)$ $D=2$ 
coefficient of the $(J)=(0+1)$ V+A correlator sum~\cite{bck04}.

\subsubsection{$D>0$ OPE Contributions}
It is the strong numerical dominance of typical OPE integrals by $D=0$ 
contributions at scales above $s_0\sim 2\ {\rm GeV}^2$ that
allows the corresponding weighted spectral integrals to 
be used in making a precision determination of $\alpha_s$. 
The impact of uncertainties in the small residual higher $D$ 
OPE terms can be understood by noting that, for all $w(s)$, 
the $D=0$ contribution to the $w(s)$-weighted OPE integral, expanded 
as a series in $a_0\equiv a(s_0)$, has the form 
$C_w\left[ 1+a_0+O(a_0^2)\right]$, where both $C_w$ and the coefficients 
occurring in the $O(a_0^2)$ contribution depend on $w(s)$. 
Since $a(m_\tau^2)\sim 0.1$, we see that a 
higher $D$ contribution with a fractional uncertainty $r$
relative to the dominant $D=0$ term will produce a
corresponding fractional uncertainty $\sim 10r$ on $\alpha_s(m_\tau^2)$.
(The factor of $10$ is reduced somewhat (to $\sim 5-6$) when
one includes the effect of higher order terms.)
Thus, e.g., to achieve a determination of $\alpha_s(M_Z^2)$ accurate
to $\sim 1\%$ (which corresponds to a determination of
$\alpha_s(m_\tau^2)$ accurate to $\sim 3\%$) one needs to 
reduce the uncertainties in the determination of the higher $D$ 
contributions, relative to the OPE total, to the sub-$0.5\%$ level.
How easy it is to satisfy this requirement depends strongly on
the choice of weight $w(s)$. We will return to this point below.

Among the $D>0$ OPE contributions, those with $D=2$ are 
either $O(m_{u,d}^2)$ or $O(\alpha_s^2 m_s^2)$~\cite{chkw93} and 
numerically negligible at the scales we consider. 
The $D=4$ OPE terms are, up to numerically tiny 
$O(m_q^4)$ corrections, determined by the RG invariant light quark, 
strange quark and gluon condensates, 
$\langle m_\ell \bar{\ell}\ell\rangle_{RGI}$,
$\langle m_s \bar{s}s\rangle_{RGI}$ and $\langle aG^2\rangle_{RGI}$.
Explicit expressions for $\left[ \Pi_{V/A}(Q^2)\right]^{OPE}_{D=4}$
may be found in Refs.~\cite{chkw93,bnp}. 

$D\geq 6$ OPE contributions
are potentially more problematic since the relevant condensates are 
either poorly known or phenomenologically undetermined. 
Defining effective condensate combinations $C_6$, $C_8$, $\cdots$ 
such that
\begin{equation}
\left[\Pi (Q^2)\right]^{OPE}_{D>4}\, \equiv\, \sum_{D=6,8,\cdots}
C_D/Q^D 
\label{dgt4opeform}\end{equation} 
up to logarithmic corrections, proportional to 
$\alpha_s log(Q^2/\mu^2)$, the $D\geq 6$ contributions 
to the RHS of Eq.~(\ref{basicfesr}), for polynomial weights,
$w(s)=\sum_{m=0} b_m y^m$, are given by
\begin{equation}
b_2\, {\frac{C_6}{s_0^2}}\, -\, b_3\, {\frac{C_8}{s_0^3}}\, +\, 
b_4\, {\frac{C_{10}}{s_0^4}}\, -\, b_5\, {\frac{C_{12}}{s_0^5}}\, 
+\, \cdots\, ,
\label{higherdintegrals}\end{equation}
again up to logarithmic corrections, proportional to 
$\alpha_s$~\cite{higherdlogsfootnote}. 
Integrated OPE contributions of $D=2k+2$ thus scale as $1/s_0^k$ 
(up to logarithms~\cite{higherdlogsdetailfootnote}), and hence as 
$1/s_0^{k+1}$ relative to the leading $D=0$ contribution.
For pinched weights, the integrals of the logarithmic corrections to
Eq.~(\ref{dgt4opeform}) are suppressed, not just by the additional 
factors of $\alpha_s$, 
but also by small numerical factors which result from the structure 
of the logarithmic integrals, 
$\oint_{\vert s\vert =s_0}ds\, y^k\, \ell n(Q^2/\mu^2)/Q^D$, and
cancellations inherent in the pinching condition $\sum_mb_m=0$. 

\subsection{The ``$(km)$ Spectral Weight'' Analyses}
Since the kinematic weight, $(1-y_\tau )^2(1+2y_\tau )$, multiplying
the $(0+1)$ spectral contribution to $R_{T;ud}$ in 
Eq.~(\ref{taukinspectral}) has degree $3$, the 
OPE representations of the $R_{T;ud}$ all contain contributions up to $D=8$,
and hence involve three unknowns, $\alpha_s$, $C^{T}_6$ and $C^{T}_8$,
which the single piece of information provided by the corresponding
total hadronic $\tau$ decay widths (or, equivalently, $R_{T;ud}$) is 
insufficient to determine.

ALEPH~\cite{dhzreview,davieretal08,alephud05,alephud9798} and 
OPAL~\cite{opalud99} dealt with this problem by constructing
additional rescaled spectral integrals, analogous to $R_{T;ud}$, 
corresponding to a range of alternate weight choices $w(s)$.
Explicitly, $\alpha_s$, $\langle aG^2\rangle_{RGI}$, 
$\delta_{V,A}^{(6)}\, =\, -24\pi^2 C^{V,A}_6/m_\tau^6$ and 
$\delta_{V,A}^{(8)}\, =\, -16\pi^2 C^{V,A}_8/m_\tau^8$ 
(or $\delta_{V+A}^{(D)}\, =\, \left(\delta_V^{(D)}+\delta_A^{(D)}\right) /2$,
with $D=6,8$) were determined as part of a combined fit
to the $s_0=m_\tau^2$ versions of the $(km)=(00),(10),(11),(12),(13)$ 
``spectral weight sum rules'', FESRs based on the weights, 
$w^{(km)}(y)=(1-y)^ky^m w^{(00)}(y)$, where 
$w^{(00)}(y)=(1-y)^2(1+2y)$ is the kinematic weight occuring on the RHS of 
Eq.~(\ref{taukinspectral}). 
ALEPH~\cite{dhzreview,davieretal08,alephud05,alephud9798} performed 
independent versions of this fit for each of the V, A and V+A channels, 
while OPAL~\cite{opalud99} performed independent fits for the V+A and 
combined V,A channels.

A crucial input to these analyses was the
assumption that $D>8$ contributions could be safely neglected
{\it for all weights considered in the fit}. In fact, 
since the polynomial coefficients relevant to $D>4$ contributions are
$(b^{(km)}_2,\cdots ,b^{(km)}_7)=(-3,2,0,0,0,0)$, $(-3,5,-2,0,0,0)$,
$(-1,-3,5,-2,0,0)$, $(1,-1,-3,5,-2,0)$ and $(0,1,-1,-3,5,-2)$
for $(km)=(00),(10),(11),(12),(13)$, respectively, we see,
from Eq.~(\ref{higherdintegrals}), that all six of the 
quantities, $C_6,\cdots , C_{16}$, would in principle contribute to 
at least one of sum rules employed, making a combined fit impossible 
without this additional assumption.

The neglect of $C_{10}$ through $C_{16}$ in the ALEPH and OPAL analyses 
creates a theoretical systematic uncertainty not included in the error 
assessments of 
Refs.~\cite{dhzreview,davieretal08,alephud05,alephud9798,opalud99}. 
Since the fits are performed with a single $s_0$ ($s_0=m_\tau^2$), 
the differing $s_0$-dependences of integrated contributions
of different $D$ are not operative, and hence neglect of
non-negligible $D>8$ contributions can be compensated for by shifts
in the values of fitted parameters relevant to lower $D$ 
contributions~\cite{bck08footnote}.
Indications that such a compensation may, indeed, be at work are provided by
(i) the lack of agreement between the values for $\langle aG^2\rangle_{RGI}$ 
obtained from the separate ALEPH V and A 
analyses~\cite{dhzreview,davieretal08}, (ii) the fact that
the central fitted values of $\langle aG^2\rangle_{RGI}$
obtained in the V, A and V+A CIPT analyses of both groups are
uniformly lower than of the updated charmonium sum rule analysis 
of Ref.~\cite{newgcond4}, and (iii) the poor quality of the 2005
ALEPH A and V+A fits ($\chi^2/dof=4.97/1$ and $3.66/1$, respectively)
and 2008 ALEPH A fit ($\chi^2/dof=3.57/1$).

A further indication that the neglect of $D>8$ contributions
(which are in principle present in the $(km)=(10),(11),(12)$ and $(13)$ 
spectral weight FESRs) is potentially dangerous is provided 
by a consideration of the relative
sizes of the $D=6,8$ and $D=0$ terms corresponding to the results
of the earlier ALEPH and OPAL fits. One should 
bear in mind that the additional factors of $y$ in
the weights $w^{(1m)}(y)$, $m\geq 1$, strongly suppress 
the correspondingly weighted $D=0$ integrals, but produce no such
suppressions of the integrated higher $D$ contributions, causing
the $D>4$ contributions to play a much larger relative role
for these weights than they do for the $(00)$ and $(10)$ weight cases. 
Taking the 2005 ALEPH V fit as an example, we find that 
\begin{itemize}
\item for the $(11)$ spectral weight FESR, the $D=6$ and $D=8$ 
contributions (which include, as per Eq.~(\ref{higherdintegrals}), the 
polynomial coefficient factors $-1$ and $-3$, respectively) represent, 
respectively, $5.2\%$ and $7.4\%$ of the leading $D=0$ contribution,
while $D=10$ and $12$ contributions (which would be weighted by 
the coefficients $5$ and $-2$ from $w^{(11)}$) are
assumed negligible;
\item for the $(12)$ spectral weight FESR, the $D=6$ and $D=8$ contributions 
(weighted by polynomial coefficients $1$ and $-1$, respectively) represent, 
respectively, $-13.7\%$ and $6.5\%$ of the $D=0$ contribution, while
$D=10,12$ and $14$ contributions (which would be accompanied by the
$w^{(12)}$ polynomial coefficients $-3$, $5$ and $-2$) are
again assumed negligible; and
\item for the $(13)$ spectral weight FESR, the $D=8$ contribution 
(weighted by polynomial coefficient $1$) represents $-14.3\%$ of the 
$D=0$ contribution, while $D=10,12,14$ and $16$ contributions (which would 
be accompanied by the $w^{(13)}$ polynomial coefficients $-1$, $-3$, $5$ 
and $-2$, respectively) are once more assumed negligible. 
\end{itemize}
Given the $< 0.5\%$ tolerance in the sum of $D>4$ relative to $D=0$
contributions required for a $\sim 1\%$ determination of 
$\alpha_s(M_Z^2)$, the neglect of $D>8$ contributions appears to us
to represent a rather strong assumption.

A quantitative test of whether or not such contributions can, in fact, 
be safely neglected {\it for all of the weights employed in the ALEPH 
and OPAL analyses} can be obtained by studying 
the quality of the fitted OPE representations of the $w^{(km)}(y)$-weighted 
spectral integrals as a function of $s_0$. The utility of this test 
follows from the fact, already noted above,
that integrated contributions of different $D$ scale differently with $s_0$. 
Thus, if the fitted values of $\alpha_s$, $\langle aG^2\rangle_{RGI}$,
$C_6$ and $C_8$ are unphysical as a result of shifts induced by the need 
to compensate for missing $D>8$ contributions in one or more of
the FESRs employed, the fact that this compensation 
occurs in lower dimension contributions, which scale more slowly with $s_0$
than do the contributions they are replacing, will show up
as a deterioration of the fit quality as $s_0$ is decreased 
below the single value $s_0=m_\tau^2$ used in the ALEPH and OPAL
analyses. In contrast, were the fit quality to be maintained at
lower $s_0$, this would provide significant evidence in support of 
the prescription of neglecting $D>8$ contributions in the set of FESRs 
employed in those analyses. We thus define the $s_0$-dependent fit-qualities,
\begin{equation}
F^w_T(s_0)\equiv {\frac{I^w_{spec}(s_0)
-I^w_{OPE}(s_0)}{\delta I_{spec}^w(s_0)}}
\label{fitqdefn}\end{equation}
where, as usual, $T=V, A$ or $V+A$, 
\begin{eqnarray}
&&I_{spec}^w(s_0)\, =\, \int_0^{s_0}ds\, w(s)\rho^{(0+1)}_{T;ud}(s)\nonumber\\
&&I_{OPE}^w(s_0)\, =\, {\frac{-1}{2\pi i}}\, \oint_{\vert s\vert =s_0}
ds\, w(s)\left[\Pi^{(0+1)}_{T;ud}(s)\right]_{OPE}
\end{eqnarray}
and $\delta I^w_{spec}(s_0)$ is the error on $I_{spec}^w(s_0)$, determined 
using the experimental covariance matrix for $dR_{T;ud}/ds$. 
One should bear in mind 
that strong correlations exist between the $I_{spec}^w(s_0)$ for fixed 
$w(s)$ but different $s_0$, and similarly between the 
$I_{OPE}^w(s_0)$ for fixed $w(s)$ but different $s_0$. 
Because of these correlations, the assumption 
that $D>8$ OPE contributions are safely negligible corresponds to the 
expectation that $\vert F^w_T(s_0)\vert$ should 
remain less than $\sim 1$ for a range of $s_0$ below $m_\tau^2$, 
{\it and for all of the $w(s)$ employed in the analysis in question}. 
It turns out that neither the ALEPH nor the OPAL fits satisfy
this expectation. 

\begin{figure}
\unitlength1cm
\caption{Fit qualities, as a function of $s_0$, for the 2005 ALEPH V fit and
the weights $w^{(00)}$, $w^{(12)}$, $w^{(13)}$, $w_2$, $w_3$ and 
$w(y)=y(1-y)^2$. The results for $w^{(00)}$, $w^{(12)}$, $w^{(13)}$, $w_2$, 
$w_3$ and $y(1-y)^2$ are shown by the dotted, medium-dashed, long-dashed, 
short dot-dashed, long dot-dashed and double-dot-dashed lines, respectively.
The right boundary corresponds to the kinematic endpoint, 
$s_0=m_\tau^2\simeq 3.16\ {\rm GeV}^2$.}
\rotatebox{270}{\mbox{
\begin{minipage}[thb]{11.2cm}
\begin{picture}(11.1,14.6)
\epsfig{figure=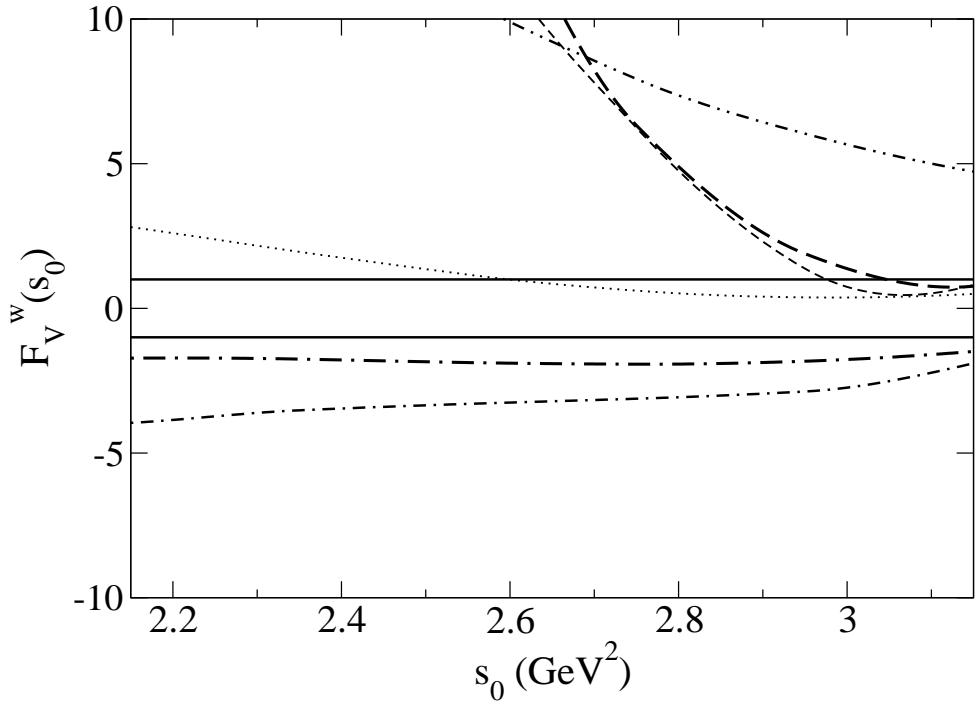,height=14.5cm,width=11.0cm}
\end{picture}
\end{minipage}}}
\label{fitqualfig}\end{figure}

To illustrate this point, we show, in Fig.~\ref{fitqualfig}, the fit 
qualities, $F_V^w(s_0)$, corresponding to the 2005 ALEPH data and 
fit~\cite{dhzreview}, for a selection of the $(km)$
spectral weights. In the figure, 
the solid horizontal lines indicate the boundaries 
$F_V(s_0)\, =\, \pm 1$ within which we would expect curves corresponding
to a physically meaningful fit to lie. We remind the reader that, although the
original 2005 ALEPH $s_0=m_\tau^2$ A and V+A fits had
$\chi^2/dof$ significantly $>1$, the $\chi^2/dof$ for the V channel
fit was $0.52/1$. The test is thus being applied to the
most successful of the previous fits. 

Also shown in the figure are the V channel fit qualities, $F_V^w(s_0)$, 
for three additional weights, $w_2(y)=(1-y)^2$, 
$w_3(y)=1-{\frac{3}{2}}y+{\frac{y^3}{2}}$ 
and $w(y)=y(1-y)^2$, all having degree $\leq 3$.
The weights $w_2$ and $w_3$ are the first two members of a series,
\begin{equation}
w_N(y)\, =\, 1\, -\, {\frac{N}{N-1}}\, y\, +\, {\frac{1}{N-1}}\, y^N
\label{wNdefn}\end{equation}
to which we will return in our own analysis below.
From Eq.~(\ref{higherdintegrals}), we see that the only $D>4$ 
contribution to the $w_2$ (respectively, $w_3$) FESR is ${\frac{C_6}{s_0^2}}$
(respectively, $-{\frac{C_8}{2s_0^3}}$). The $w_2$ (respectively $w_3$) FESR
thus provides a useful independent test of the value
of $C_6$ (respectively $C_8$) obtained in the earlier fits. 
The $w(y)=y(1-y)^2$ FESR, with $D>4$ OPE contribution
$-{\frac{2C_6}{s_0^2}}\, -\, {\frac{C_8}{s_0^3}}$,
provides another such test since this linear combination is independent of that
appearing in the $(00)$ spectral weight FESR. The strength of the test
is enhanced in this case because the factor $y$ in the weight 
leads to a significant suppression of the $D=0$ integral, making the 
$y(1-y)^2$ FESR relatively more sensitive to $D>4$ contributions. 
If the neglect of $D>8$ contributions in the earlier analyses
was actually justified, the $s_0<m_\tau^2$ FESRs corresponding not only to 
the spectral weights employed in those fits, but also to $w_2$, $w_3$, 
and $y(1-y)^2$ should all be well-satisfied using the 
fitted values of the input $D\leq 8$ OPE 
parameters. It is evident from the figure that this is far from being the
case. The poor quality of the ALEPH fit when applied to the $w_2$, $w_3$ and
$y(1-y)^2$ FESRs, even at $s_0=m_\tau^2$, 
and the fact that the nominally good quality of the 
original fit to the $s_0=m_\tau^2$ spectral weight FESRs 
does not persist to lower $s_0$, clearly establish the presence of $D>8$ 
contamination in at least some of the original fitted FESRs. 
The deterioration in the fit quality as $s_0$ is decreased below
$m_\tau^2$ seen for all cases shown in the figure 
is in fact a general feature, one found for all of the weights discussed 
and all three of the channels investigated in this paper.


One could, of course, attempt to use the $s_0$ dependence of the 
$w^{(km)}$-weighted spectral integrals to aid in
achieving an improved fit for the $D>4$ $C_D$. 
It is important to bear in mind, however, that the range of 
$s_0$ that can be employed in such a fit is limited:
to $s_0<m_\tau^2$ by kinematics, and to $s_0$ greater than 
$\sim 2\ {\rm GeV}^2$, if one wishes to avoid non-negligible
``duality violation'' (OPE breakdown)~\cite{gik01,cdgmvma,cgp08}. In 
such a relatively restricted window, the number of independent parameters that
can be successfully fitted is limited. The $(km)$ spectral weight
FESRs thus represent non-optimal choices for an analysis of this type
since their OPE sides typically involve, in addition to the 
parameter $\alpha_s(m_\tau^2)$ we are primarily interested in determining, a 
combination of several of the unknown $D>4$ $C_D$. It is also
worth stressing that the $(11)$, $(12)$ and $(13)$ spectral weight FESRs used 
in the previous analyses have another feature which makes
them non-optimal for an analysis whose main goal is the determination 
of $\alpha_s$. Optimization of such a determination is achieved by
using sum rules which enhance, as much as possible, the
relative contribution of the integrated $D=0$ series, since it is in
this contribution that the dominant dependence on $\alpha_s$ lies.
The $(1m)$, $m\geq 1$, spectral weights, however, do exactly the opposite, 
the additional factors of $y$ producing rather strong suppressions
of the leading $D=0$ OPE integrals (by factors of $\sim$ $6.5,\ 17$,
and $37$ relative to the corresponding $(00)$ integral for the $(11)$,
$(12)$ and $(13)$ cases, respectively) without any accompanying 
suppression of higher $D$ contributions (beyond that which may (or may not)
be present in the correlator itself).

\subsection{An Alternate Analysis Strategy}

In view of the problems displayed by the $(km)$ spectral weight FESR 
analyses, we turn to FESRs based on the weights, 
$w_N(y)$ introduced already in Eq.~(\ref{wNdefn}) above. 
The $w_N$ are constructed to share with the $(00)$ spectral
weight the presence of a double zero at $s=s_0$ and the resulting
suppression of OPE-violating contributions near the timelike
point on the OPE contour. For our problem they have, in addition,
the following positive features, not shared by the 
set of $(km)$ spectral weights employed in the ALEPH and OPAL
analyses:
\begin{itemize}
\item the $D=0$ integrals grow moderately with $N$ rather than decreasing
strongly as was the case when one went from the lower to the higher
spectral weights; 
\item at the same time, the coefficient governing
the only unsuppressed $D>4$ contribution (that with $D=2N+2$)
decreases with $N$, further enhancing $D=0$ relative to $D>4$
contributions; 
\item because each $w_N$ FESR involves only a single unsuppressed 
$D>4$ contribution, the collection of $w_N$ FESRs is well-adapted to 
most efficiently implementing the constraints associated with the $s_0$ 
dependence of the correspondingly weighted spectral integrals 
in the fitting of the unknown $D>4$ OPE parameters; and
\item as $N$ is increased, the $1/s_0^{N+1}$ scaling of the single
unsuppressed $D=2N+2$ contribution relative to the leading $D=0$
contribution varies more and more strongly with $s_0$, increasing the
leverage for fitting $C_{2N+2}$ (though the effect is of course offset
to some extent by the decrease with $N$ of the polynomial
coefficient, $1/(N-1)$, present in the integrated form of the $D=2N+2$
contribution).
\end{itemize}

To quantify the extent to which the level of $D=0$ dominance of the $w_N$ 
FESRs represents an improvement over that of the $(km)$ spectral weight 
FESRs, we introduce the double ratio, $R^D[w_N,w^{(km)},s_0]$,
defined by
\begin{equation}
R^D[w_N,w^{(km)},s_0]\ =\ {\frac{r_{w_N}^D(s_0)}{r_{w^{(km)}}^D(s_0)}}
\label{dblratio}\end{equation}
where 
\begin{equation}
r_w^M(s_0) \equiv {\frac{\left[ I_{OPE}^{w}(s_0)\right]_{D=M}}
{\left[ I_{OPE}^{w}(s_0)\right]_{D=0}}}\ .
\label{basicratio}\end{equation}
$R^D[w_N,w_{km},s_0]$ represents the suppression of the
fractional contribution of dimension $D$ in the $w_N$ FESR relative
to that in the $w^{(km)}$ FESR and, by construction, is independent of $C_D$.
Taking $s_0=m_\tau^2$ to be specific, we find that
\begin{itemize}
\item $R^6[w_2,w^{(km)},m_\tau^2]\, =\, -1/2.1,\, -1/2.9,\, -1/4.4,$ and 
$-1/12$ for $(km)=(00)$, $(10)$, $(11)$ and $(12)$, respectively;
\item $R^8[w_3,w^{(km)},m_\tau^2]\, =\, 1/3.1,\, 1/11,\, -1/25$,
$-1/26$ and $-1/58$ for $(km)=(00)$, $(10)$, $(11)$, $(12)$ and $(13)$, 
respectively;
\item $R^{10}[w_4,w^{(km)},m_\tau^2]\, =\, -1/6.8,\, 1/79,\, -1/126$,
and $-1/91$ for $(km)=(10)$, $(11)$, $(12)$ and $(13)$, 
respectively;
\item $R^{12}[w_5,w^{(km)},m_\tau^2]\, =\, -1/44,\, 1/288$
and $-1/379$ for $(km)=(11)$, $(12)$ and $(13)$, respectively; and
\item $R^{14}[w_6,w^{(km)},m_\tau^2]\, =\, -1/149$ and $1/814$ 
for $(km)=(12)$ and $(13)$, respectively.
\end{itemize}
Neglect of $D>8$ contributions would thus be 
between $\sim$ $1$ and $3$ {\it orders of magnitude} safer for the
$w_4$, $w_5$ and $w_6$ FESRs than it would for the $(10),(11),(12)$
and $(13)$ spectral weight sum rules. Had it been safe for the latter, then
it would certainly also be safe for the former. From our
fits below, however, we find small, but not entirely negligible, $D=10,12,14$
contributions to the $w_4$, $w_5$ and $w_6$ FESRs, respectively.
The analogous contributions, which play a much larger relative role 
in the higher spectral weight FESRs, account for the problems 
of the ALEPH and OPAL spectral weight FESR fits seen in the fit
quality plot above.

\section{\label{sec3}The $w_N$ FESR analyses}

As $N$ gets large, the different $w_N(y)$ become less and less independent,
approaching $1-y$ in the limit that $N\rightarrow\infty$.
The approach to $1-y$ also weakens the level of the desired suppression
of contributions from the vicinity of the timelike point on the OPE contour.
In addition, the reduction of the unsuppressed integrated $D=2N+2$ contribution
by the factor $1/(N-1)$ means that these contributions will eventually be
driven down to the level of the other, numerically and $\alpha_s$-suppressed,
contributions of $D>4$ having $D\not= 2N+2$~\cite{wrongdlogsfootnote}. 
For these reasons we focus, in what follows,
on those FESRs corresponding to the limited set of weights 
$w_2,\cdots ,w_6$. A clear demonstration of the independence of the 
results associated with the different $w_N$ in this set will be given
in Section~\ref{sec4}.

The values of any input parameters, together with details of our treatment of
the spectral and OPE integral sides of the $w_N$ FESRs, are given in 
Subsections~\ref{subsecspec} and \ref{subsecope}, respectively.
Results for the ALEPH-based V, A and V+A and OPAL-based V+A
fits, as well as a breakdown of the contributions to the theoretical
errors on the fitted parameters, $\alpha_s(m_\tau^2)$ and
$C_D$, $D=6,\, 8\cdots 14$, are given in subsection~\ref{subsecresults}.
A final assessment and discussion of the results is deferred to
Section~\ref{sec4}.

\subsection{\label{subsecspec}The $w_N$-weighted spectral integrals}
On the spectral integral side of the $w_N$ FESRs, we employ for our main 
analysis the publicly available
2005 ALEPH V, A and V+A spectral data and covariance
matrices~\cite{dhzreview,alephud05}.
Our central results will also follow Ref.~\cite{davieretal08}
in incorporating, in the V and A channels, the improved $s$-dependent 
V/A separation of the contribution from the $K\bar{K}\pi$ mode made possible 
by the recent BaBar isovector electroproduction 
cross-section measurements~\cite{babarkkbarpi07} and the details
on the $V+A$ $K\bar{K}\pi$ distribution presented in 
Ref.~\cite{davieretal08}. Independent analyses using the
1999 OPAL V, A and V+A data and covariance matrices have also been
performed, though in this case we do not have the information
on the $K\bar{K}\pi$ distribution needed to make the improved
V/A separation for that mode and so will report results below
only for the V+A analysis.

We employ as input to the determination of the isovector spectral
function from the ALEPH or OPAL distributions the values
\begin{eqnarray}
S_{EW}&&=1.0201(3)\\
B_e&&=0.17818(32)\\
\vert V_{ud}\vert &&= 0.97408(26)
\end{eqnarray}
where $S_{EW}$ is taken from Ref.~\cite{erler}, the 
lepton-universality-constrained result for $B_e$ from
Ref.~\cite{banerjee07}, and the result for $\vert V_{ud} \vert$
from the most recent update of the $0^+\rightarrow 0^+$ superallowed nuclear 
$\beta$ decay analysis~\cite{hardydec07}. The
$\pi$ pole contribution to the A and V+A spectral integrals 
is evaluated using the very accurate determination of
$f_\pi \vert V_{ud}\vert$ from the $\pi_{\mu 2}$ width~\cite{pdg06}. A small 
global renormalization must also be applied to the ALEPH
and OPAL data as a result of small changes to 
$B_e$, $S_{EW}$, $\vert V_{ud}\vert$ and the total $\tau$ strange 
branching fraction, $B_s$, (which enters the most precise determination of 
the overall V+A normalization, $R_{ud;V+A}$) since the original publications.
With the full set of recent BaBar and Belle updates to the branching 
fractions of various strange modes~\cite{strangeBvals}, we obtain 
$R_{ud;V+A}=3.478(11)$. It is assumed that the
continuum parts of the V, A and V+A distributions 
are all to be rescaled by the same common factor. The uncertainty in 
$R_{ud;V+A}$ strongly dominates the overall normalization uncertainty
on the spectral integrals.

\subsection{\label{subsecope}The $w_N$-weighted OPE integrals}

For the $D=0$ contribution we employ the CIPT evaluation as our central 
determination. We truncate the $D=0$ Adler function series at $O(\bar{a}^5)$, 
using the known coefficients for terms up to $O(\bar{a}^4)$ and the 
estimate $d^{(0)}_5=275\pm 275$ of Ref.~\cite{bck08} for the coefficient 
of the last term. An independent evaluation using the alternate
FOPT evaluation is also performed and the variation induced by
the uncertainty in $d^{(0)}_5$ and the CIPT-FOPT difference added in
quadrature to produce the full truncation uncertainty estimate. An
analogous procedure, using however the average of the CIPT and
FOPT determinations as central value, and half the difference
as the corresponding component of the truncation uncertainty
estimate (added linearly to the uncertainty generated by that
on $d^{(0)}_5$), was employed in Ref.~\cite{bck08}. Our estimate 
yields a $D=0$ truncation uncertainty 
assessment similar to that of Ref.~\cite{bck08},
but significantly more conservative than the alternate estimates based on
a combination of the $d^{(0)}_5$ uncertainty and residual scale dependence
which have also been employed elsewhere in the literature. 

In evaluating the running coupling over the OPE contour we employ 
the exact analytic solution associated with the $4$-loop-truncated 
$\beta$ function~\cite{4loopbeta}. The reference scale input
needed to specify this solution, taken here to be $\alpha_s(m_\tau^2)$,
is to be determined as part of the fitting procedure.

The $D=2$ contributions, as already noted, are either $O([m_d\pm m_u]^2)$ 
or $O(\alpha_s^2m_s^2)$, and hence expected to be numerically negligible. 
Our central values correspond to neglecting them entirely. The 
$O([m_d\pm m_u]^2)$ contributions should, in fact, be neglected in 
any case, as a matter of consistency. The reason is that, even at 
the highest scale, $s_0=m_\tau^2$, allowed by kinematics, the OPE 
representation of the ``longitudinal'' ($J=0$) contribution to the 
experimental spectral distribution (in the $(J)=(0+1)/(0)$ decomposition of 
Eq.~(\ref{taukinspectral})) is completely out of control. Not only do the 
variously weighted integrated $D=2$ OPE series display extremely bad 
convergence, but all truncation schemes for these badly behaved series 
employed in the literature badly violate constraints associated with 
spectral positivity~\cite{kmlongproblem}. It is thus impossible to use the
longitudinal OPE to estimate the $O([m_d\pm m_u]^2)$ longitudinal
contributions to the spectral distribution, which means that the
spectral functions $\rho^{(0+1)}_{ud;V/A}(s)$ can be determined
only up to uncertainties of $O([m_d\mp m_u]^2)$, respectively. It would thus 
be inconsistent to explicitly include contributions of this same
order on the OPE side of the $0+1$ FESRs. We have, in any case,
verified, by direct computation, that including the integrated $J=0+1$,
$D=2$ OPE contributions would have a negligible impact on our analysis,
in agreement with the results for these contributions quoted in the earlier 
analyses. The $J=0+1$, $D=2$ computation employed the exact solution for 
the running masses corresponding to the $4$-loop truncated 
$\beta$~\cite{4loopbeta} and $\gamma$~\cite{4loopgamma} functions, with
PDG06 values for the $\overline{MS}$ scheme light and
strange quark masses at scale $2$ GeV~\cite{pdg06} as input. 
It is also possible to
estimate the contributions from the non-$\pi$-pole part of the $J=0$
spectral distributions and verify that they are safely negligible.
For the A channel this estimate employs the spectral
model of Ref.~\cite{kmjkudps} for the isovector pseudoscalar channel,
a model generated using a combined Borel and finite energy sum rule analysis
of the relevant pseudoscalar correlator~\cite{kmjkudps}. 
The isovector V channel $J=0$ contributions,
being suppressed by a further factor of $[(m_d-m_u)/(m_d+m_u)]^2\sim 1/10$
are even more negligible.

We employ as basic $D=4$ input 
\begin{eqnarray}
&&\langle 2{m_\ell}\bar{\ell}\ell\rangle_{RGI}\, =\,  -m_\pi^2f_\pi^2
\ {\rm and}\\
&&\langle aG^2\rangle_{RGI}\, =\, (0.009\pm 0.007)\ {\rm GeV}^4
\label{d4input}\end{eqnarray}
the first result being the GMOR relation~\cite{gmor}
and the second the result of Ref.~\cite{newgcond4}.
The remaining $D=4$ combination,
$\langle m_s\bar{s}s\rangle_{RGI}$, then follows from
conventional ChPT quark mass ratios~\cite{leutwylermq} and the value,
\begin{equation}
r_c={\frac{\langle \bar{s}s\rangle_{RGI}}{\langle \bar{\ell}\ell\rangle_{RGI}}}
=1.1\pm 0.6\ ,
\end{equation}
obtained by updating the analysis of 
Ref.~\cite{jaminssoverll}, using the range of recent $n_f=2+1$ lattice results
for $f_{B_s}/f_B$ as input~\cite{latticefbsoverfb}.
Although this value of $r_c$ is nearly twice that employed in the earlier
ALEPH and OPAL analyses (whose values, however, are based on somewhat 
out-of-date input), the difference between the two has negligible impact 
on the final analysis since the integrated $D=4$ contributions are both 
small at the scales employed
and, in any case, dominated by the gluon condensate contribution. 
The sizable uncertainty we quote on $r_c$, for the same reason,
plays a negligible role in our final theoretical error estimate.
%

$D>4$ contributions are handled by treating the various $C_{2N+2}$ as 
fit parameters. $C_{2N+2}$ is fitted, together with 
$\alpha_s(m_\tau^2)$, to the set of $I_{w_N}(s_0)$ 
corresponding to a range of $s_0$. The requirement that the
values of $\alpha_s(m_\tau^2)$ obtained in this manner from
the different $w_N$ FESRs should be consistent provides 
a non-trivial check on the reliability of the analysis.
We discuss this issue further in Section~\ref{sec4}.

For the ALEPH-based fits, we work with an equally spaced set of $s_0$ values,
$s_0=(2.15\, +\, 0.2k)\ {\rm GeV^2}$, $k=1,\cdots , 6$, adapted to the ALEPH 
experimental bins. We also study the stability of our fits by either
removing the $2.15\ {\rm GeV}^2$ point or adding, in addition,
$s_0=1.95\ {\rm GeV}^2$. For the OPAL-based fits, the analogous
$s_0$ set is $s_0=(2.176\, +\, 0.192k)\, {\rm GeV}^2$, $k=1,\cdots , 6$,
with stability studied by either removing the lowest point, or
adding an additional point with $s_0=1.984\ {\rm GeV}^2$.

\subsection{\label{subsecresults}Results}

Results for the V, A and V+A fits based on the ALEPH data are presented
in the upper portion of Table~\ref{table1}. In the table, we display,
for each of the $w_N$, $N=2,\cdots ,6$, FESRs, the fitted values of  
$\alpha_s(m_\tau^2)$ and the relevant $D>4$ coefficient, $C_{2N+2}$, the 
latter quoted in the dimensionless form, $C_{2N+2}/m_\tau^{2N+2}$.
We remind the reader that, in arriving at these values,
we have implemented the improved V/A separation 
for the $K\bar{K}\pi$ mode, discussed already above. This improvement 
produces an upward (downward) shift of $0.0013$ in the central value 
of the A (V) determinations of $\alpha_s(m_\tau^2)$, improving further 
the consistency between the results of the separate V, A and V+A analyses. 
The level of consistency, even before this improvement, 
is significantly better than that displayed by the 
$(km)$ spectral weight analysis results reported in Ref.~\cite{davieretal08}.

The lower portion of Table~\ref{table1} contains the corresponding results for 
the OPAL-based V+A fits. The results for the separate V and A fits are
not displayed in this case, since we lack the information 
on the $K\bar{K}\pi$ contribution to the inclusive distribution required to 
perform the improved V/A separation. For completeness, however,
we mention that the central values of $\alpha_s(m_\tau^2)$ obtained 
without this correction lie
$0.003$ lower (higher) for the V (A) fits. The improved V/A separation,
of course, plays no role in the V+A fit. The ALEPH- and OPAL-based
results are seen to be in very good agreement within errors.

\begin{table}
\caption{\label{table1}Results of the $w_N$ FESR fits for 
$\alpha_s(m_\tau^2)$ and $C_{2N+2}/m_\tau^{2N+2}$ obtained 
using either the ALEPH or OPAL data and covariances. In all
entries, the first error is experimental and the second theoretical.}
\vskip .1in
\begin{tabular}{|c|c|c|c|l|}
\hline
Data set&Channel&Weight $w_N$&
$\alpha_s\left( m_\tau^2\right)$&\qquad $C_{2N+2}/m_\tau^{2N+2}$\\
\hline
ALEPH&V&$w_2$&\ \ $0.321(7)(8)$\ \ &\ \ $-0.000187(29)(56)$\ \ \ \\
&&$w_3$&\ \ $0.321(7)(10)$\ &$\ \ \ \ 0.000060(36)(60)$\ \ \ \\
&&$w_4$&\ \ $0.321(7)(11)$\ &$\ \ \ \ 0.000015(36)(53)$\ \ \ \\
&&$w_5$&\ \ $0.321(7)(12)$\ &\ \ $-0.000043(33)(44)$\ \ \ \\
&&$w_6$&\ \ $0.321(7)(12)$\ &$\ \ \ \ 0.000046(27)(35)$\ \ \ \\
\hline
&A&$w_2$&\ \ $0.319(6)(9)$\ \ &\ \ $-0.000072(24)(60)$\ \ \ \\
&&$w_3$&\ \ $0.319(6)(10)$\ &$\ \ \ \ 0.000182(28)(71)$\ \ \ \\
&&$w_4$&\ \ $0.319(6)(11)$\ &\ \ $-0.000216(27)(70)$\ \ \ \\
&&$w_5$&\ \ $0.319(6)(12)$\ &\ \ $\ \ 0.000201(23)(66)$\ \ \ \\
&&$w_6$&\ \ $0.319(6)(12)$\ &\ \ $-0.000166(19)(59)$\ \ \ \\
\hline
&V+A&$w_2$&\ \ $0.320(5)(8)$\ \ &\ \ $-0.000261(35)(114)$\ \ \\
&&$w_3$&\ \ $0.320(5)(9)$\ \ &$\ \ \ \ 0.000247(45)(125)$\ \ \\
&&$w_4$&\ \ $0.320(5)(10)$\ &\ \ $-0.000208(44)(111)$\ \ \\
&&$w_5$&\ \ $0.320(5)(11)$\ &$\ \ \ \ 0.000166(39)(97)$\ \ \ \\
&&$w_6$&\ \ $0.320(5)(12)$\ &\ \ $-0.000126(34)(88)$\ \ \ \\
\hline
\hline
OPAL&V+A&$w_2$&\ \ $0.322(7)(8)\ \ $&\ \ $-0.000233(59)(114)$\ \ \\
&&$w_3$&\ \ $0.322(7)(10)$\ &$\ \ \ \ 0.000205(74)(120)$\ \ \\
&&$w_4$&\ \ $0.322(7)(11)$\ &\ \ $-0.000162(76)(105)$\ \ \\
&&$w_5$&\ \ $0.322(7)(12)$\ &$\ \ \ \ 0.000122(70)(86)$\ \ \ \\
&&$w_6$&\ \ $0.322(8)(12)$\ &\ \ $-0.000091(60)(67)$\ \ \ \\
\hline
\end{tabular}
\end{table}

The experimental errors quoted in the table contain a component 
associated with the $0.32\%$ normalization uncertainty, which is
$100\%$ correlated for all of the separate analyses. The theory error 
is obtained by adding in quadrature uncertainties associated
with (i) the truncation of the $D=0$ series (itself the quadrature
sum of the difference of the CIPT and FOPT fit results
and the uncertainty produced by taking $d^{(0)}_5=275\pm 275$), (ii) the
uncertainties on the $D=4$ input condensates and (iii) the
``stability'' uncertainty, generated by varying the lower 
edge of the fit window employed, as described above. 

Individual contributions to the theoretical errors on the fitted
parameters, $\alpha_s(m_\tau^2)$ and $C_{2N+2}/m_\tau^{2N+2}$,
obtained from the $w_N$-weighted, ALEPH-based V+A FESRs, are shown, 
in the upper and lower halves of Table~\ref{therrtable}, respectively.
Results for the OPAL-based V+A and ALEPH-based V and A fits are not 
quoted separately, the decompositions being similar, with the exception of the
stability contributions for the OPAL-based V+A fits, which are a factor of 
$\sim 2$ smaller than those for the corresponding ALEPH-based V+A fits.
The differences between the results produced by the CIPT and FOPT 
evaluations of the $D=0$ OPE contributions are given in the $FOPT$
column of the table, while the uncertainties associated with those
on $d^{(0)}_5$, $\langle aG^2\rangle_{RGI}$, and the variation
of the lower edge of the $s_0$ fit window appear in the columns
headed by $\delta d^{(0)}_5$, $\delta \langle aG^2\rangle$,
and $stability$, respectively. The very small uncertainties generated by those
on the light and strange condensates (which, for example, produce
uncertainties of $\pm 0.0002$ on $\alpha_s(m_\tau^2)$) can be neglected 
without changing the total theoretical error, and hence are not quoted 
explicitly in the table. In all cases we symmetrize the quoted errors, 
taking the larger of the two possibilities in the event that the original 
error is asymmetric.

\begin{table}
\caption{\label{therrtable}Contributions to the theoretical
uncertainties on $\alpha_s(m_\tau^2)$ and $C_{2N+2}/m_\tau^{2N+2}$
obtained in the fits to $w_N$ V+A FESRs based on the ALEPH data
and covariances.}
\vskip .1in
\begin{tabular}{|l|c|cccc|}
\hline
Observable&Weight $w_N$
&\ \ \ \  $FOPT$\ \ \ \ &\ \ \ \ $\delta d^{(0)}_5$\ \ \ \ 
&\ \ \ \ $\delta\langle aG^2\rangle$\ \ \ \ &\ \ \ $stability$\ \ \ \\
\hline
$\alpha_s(m_\tau^2)$&$w_2$&$0.0004$&$0.0056$&$0.0059$&$0.0014$\\
&$w_3$&$0.0049$&$0.0056$&$0.0059$&$0.0014$\\
&$w_4$&$0.0068$&$0.0056$&$0.0059$&$0.0013$\\
&$w_5$&$0.0079$&$0.0055$&$0.0059$&$0.0013$\\
&$w_6$&$0.0084$&$0.0056$&$0.0059$&$0.0015$\\
\hline
$C_{2N+2}/m_\tau^{2N+2}$
&$w_2$&$0.000069$&$0.000019$&$0.000084$&$0.000027$\\
&$w_3$&$0.000090$&$0.000016$&$0.000072$&$0.000044$\\
&$w_4$&$0.000078$&$0.000013$&$0.000058$&$0.000053$\\
&$w_5$&$0.000063$&$0.000012$&$0.000045$&$0.000058$\\
&$w_6$&$0.000051$&$0.000008$&$0.000035$&$0.000062$\\
\hline
\end{tabular}
\end{table}

We see from the table that the contributions to the theoretical error 
on $\alpha_s(m_\tau^2)$ are very similar for the various $w_N$,
with the exception of the FOPT-CIPT difference, which is small
for $w_2$ and grows with increasing $N$. One should bear in mind, however,
that, for the kinematic weight, $w^{(00)}$, the FOPT expansion,
truncated at a given order, was shown to oscillate about the
correspondingly truncated CIPT expansion with a period of about
$6$ perturbative orders~\cite{jaminfopt05}. Studying the FOPT-CIPT
difference as a function of truncation order for the various
$w_N$ we find evidence for a similar oscillatory pattern, but with the
truncation order at which the cross-over between the two truncated sums 
occurs dependent on $N$. We thus consider the small FOPT-CIPT difference
for $w_2$ an artifact of the particular truncation order of
our central results, and expect the difference to grow
for the next few truncation orders. For this reason, to be conservative, we 
take the largest of the FOPT-CIPT differences (that for $w_6$) 
as our estimate of the FOPT vs. CIPT component of the truncation uncertainty
for $\alpha_s(m_\tau^2)$ for all of the $w_N$ FESRs studied. This
prescription leads to a common theoretical error of 
$\pm 0.012$ for all of our determinations of $\alpha_s(m_\tau^2)$.

The results quoted so far take into account short-distance electroweak
corrections but do not include long-distance electromagnetic (LDEM) effects.
Such LDEM corrections, though believed to be small, have
been investigated in detail only for the $\pi\pi$ final hadronic 
state~\cite{cen,ffls06}. We study the impact of the $\pi\pi$ LDEM 
corrections on the V and V+A channel analyses using the form of these 
corrections given in Ref.~\cite{ffls06} (which implementation 
incorporates a resonance contribution not included 
in the earlier studies of Refs.~\cite{cen}). We find that the
correction raises $\alpha_s(m_\tau^2)$ by $0.0002-0.0003$ 
($0.0001-0.0002$) for the 
various V (V+A) channel $w_N$ FESR analyses. In arriving
at our final assessment, reported in the next section, we have included 
the $\pi\pi$ LDEM correction, assigning it an uncertainty of $100\%$, in
view of the as-yet-undetermined corrections associated with higher multiplicity
modes. Even were one to expand this uncertainty several-fold,
the impact on our final error would remain entirely negligible.

\section{\label{sec4}Discussion and final results}
\subsection{\label{subsecdiscussion}Discussion}
In this subsection we discuss further the reliability and consistency
of our extraction of $\alpha_s$, compare our results for the $C_D$ with those 
of other analyses, and comment on a number of other relevant points.

\subsubsection{Impact of the new Belle $\pi\pi$ data}
We begin by discussing what impact the recently released Belle
$\tau\rightarrow \pi\pi\nu_\tau$ data~\cite{bellepipi} 
might have on our conclusions.
Note that the $\pi\pi$ branching fraction, $B_{\pi\pi}$, measured by Belle
is in good agreement with the previous $\tau$ 
measurements reported by ALEPH~\cite{alephud05}, OPAL~\cite{opalud99},
CLEO~\cite{cleoud94}, L3~\cite{l3ud95} and DELPHI~\cite{delphiud06}.
The unit-normalized number distribution, however, differs slightly in shape
from that obtained by ALEPH, being somewhat higher (lower) than
ALEPH below (above) the $\rho$ peak. Such a difference will lead
to normalization and $s_0$-dependence shifts in the weighted
V and V+A spectral integrals, causing, in general, shifts in the fitted
values of both $\alpha_s(m_\tau^2)$ and the $C_{2N+2}$.
To investigate the size of these effects,
we use the new world average for $B_{\pi\pi}$ (including the
Belle result) to fix the overall normalization of the Belle $\pi\pi$
distribution and, after adding the difference of the weighted BELLE and
ALEPH $\pi\pi$ spectral integral components to the ALEPH spectral integrals,
perform a series of ``Belle-$\pi\pi$-modified'' $w_N$ FESR fits. 
Since we lack the covariance information needed to fully replace the
ALEPH $\pi\pi$ with Belle $\pi\pi$ data, we employ the
ALEPH covariance matrix, without change, in the fit. The results
thus represent only an exploration of the magnitude of the shift
in $\alpha_s$ likely to be associated with such a shift in the
shape of the $\pi\pi$ distribution. We find that the 
Belle-$\pi\pi$-modified V channel (respectively, V+A channel) fits 
yield $\alpha_s(M_Z^2)$ values lower than those obtained using the
ALEPH data alone by $\sim 0.00007$ (respectively, $0.00013$), showing that
the impact on our central result (obtained from the V+A channel fits)
is negligible on the scale of our other uncertainties. It would nonetheless 
be extremely interesting to have measured versions of the full non-strange 
spectral distribution, including the improved V/A separation made possible
by the much higher statistics, from the B factory experiments.

\subsubsection{Consistency and reliability of the analysis}
With regard to the reliability and consistency of our results, we note first
that, for each of the V, A and V+A analyses, the same quantity, 
$\alpha_s(m_\tau^2)$, is obtained from five independent FESR fits.
In each of the V, A and V+A channels, we find that the results from 
the different $w_N$ analyses are in exceedingly good
agreement, the variation across the different weight choices being
at the $\pm 0.0001$ level, and hence invisible at the precision
displayed in Table~\ref{table1}. The fitting of the $D>4$
OPE coefficients, $C_D$, and concommitant identification of the small
$D>4$ OPE contributions is crucial to achieving this level of agreement,
as can be seen from Table~\ref{table3}, which shows the ALEPH V+A fit
values for $\alpha_s(m_\tau^2)$ already quoted above, together with 
the corresponding results obtained by ignoring the relevant
$D>4$ contribution, and working at the highest available scale,
$s_0=m_\tau^2$. In assessing the improvement in consistency
produced by including the $C_D$ in the fits, one should bear in
mind that the non-normalization component of the experimental
uncertainty (which is still correlated but, unlike the normalization 
and theoretical uncertainties, not $100\%$ correlated amongst the different 
weight cases) is $0.003$. The impact of including the $D>4$ contributions
is, not surprisingly, greatest for the $w_2$ FESR, where the suppression
of the $D=6$ contribution by the polynomial coefficient factor $1/(N-1)$
($=1$ in this case) is the least strong of all the cases studied. The 
results of the table also show that use of the $w_N$ FESRs has
(as intended) been successful in suppressing $D>4$ relative to $D=0$ 
OPE contributions, an effect desirable for optimizing
the accuracy of our $\alpha_s$ determination. The table in fact shows that
the impact of the full $D>4$ contribution, in all but the $w_2$ case, 
is at a level less than $\sim 50\%$ of the dominant theoretical component
of the overall uncertainty, making the
impact of higher order corrections to the treatment of the
integrated $D>4$ contributions safely
negligible~\cite{higherdlogsdetailfootnote}.

\begin{table}
\caption{\label{table3}Impact of the inclusion of $D>4$ OPE contributions
on the fitted values for $\alpha_s(m_\tau^2)$ for the ALEPH-based
analyses. The column headed $full\ fit$ repeats the values quoted
above for the various $w_N$-weighted V+A FESRs, while that headed 
$no\ D>4$ contains the corresponding values obtained
by working at the maximum scale $s_0=m_\tau^2$ and neglecting
the contribution of dimension $D=2N+2$ on the OPE side.}
\vskip .1in
\begin{tabular}{|l|c|cc|}
\hline
Channel&Weight&\ \ \ \ $full\ fit$\ \ \ \ &\ \ \ $no\ D>4$\ \ \ \ \\
\hline
V&$w_2$&$0.321$&$0.305$\\
&$w_3$&$0.321$&$0.320$\\
&$w_4$&$0.321$&$0.323$\\
&$w_5$&$0.321$&$0.325$\\
&$w_6$&$0.321$&$0.325$\\
\hline
A&$w_2$&$0.319$&$0.314$\\
&$w_3$&$0.319$&$0.312$\\
&$w_4$&$0.319$&$0.314$\\
&$w_5$&$0.319$&$0.316$\\
&$w_6$&$0.319$&$0.318$\\
\hline
V+A&$w_2$&$0.320$&$0.310$\\
&$w_3$&$0.320$&$0.316$\\
&$w_4$&$0.320$&$0.319$\\
&$w_5$&$0.320$&$0.321$\\
&$w_6$&$0.320$&$0.322$\\
\hline
\end{tabular}
\end{table}

While the lack of consistency of the results for $\alpha_s$ in the
limit that all the $C_D$ are set to zero establishes the independence 
of the different $w_N$-weighted FESRs, and hence the non-trivial
nature of the consistency observed once the $C_D$ are included in the fits,
an even more compelling case for the degree of independence of the
different FESRs is provided by the results obtained by fitting the 
$w_N$-weighted OPE integrals to the set of $w_M$-weighted spectral integrals,
with $N\not= M$.
The results for $\alpha_s(m_\tau^2)$ obtained 
from this exercise, using the ALEPH data in the V+A channel, are shown 
in Table~\ref{indeptable}, whose
row (respectively, column) headings give the weight employed for the spectral 
(respectively, OPE) integrals. Blank entries in the table denote
cases where no minimum could be found for the $\chi^2$ function 
having positive $\alpha_s(m_\tau^2)$. It is evident from the table 
that the constraints on $\alpha_s$ associated with the set of 
$w_N$ employed in our analysis enjoy a high degree of independence.

\begin{table}
\caption{\label{indeptable}The fitted values for $\alpha_s(m_\tau^2)$ 
obtained from an ALEPH-based V+A analysis employing one $w_N$ for the 
spectral integrals (identified by the row label) but a different 
$w_N$ for the OPE integrals (identified by the column heading).}
\vskip .1in
\begin{tabular}{|c|ccccc|}
\hline
&$w_2$&$w_3$&$w_4$&$w_5$&$w_6$\\
\hline
$w_2$&$0.320$&$0.175$&---&---&---\\
$w_3$&$0.435$&$0.320$&$0.249$&$0.194$&$0.149$\\
$w_4$&$0.499$&$0.384$&$0.320$&$0.277$&$0.243$\\
$w_5$&$0.541$&$0.423$&$0.361$&$0.320$&$0.291$\\
$w_6$&---&$0.450$&$0.388$&$0.349$&$0.320$\\
\hline
\end{tabular}
\end{table}

\begin{figure}
\unitlength1cm
\caption{Comparison of the fit qualities corresponding to (i) our fits
and (ii) the 2005 ALEPH fit, as a function of $s_0$, for the V channel and 
the weights $w^{(00)}$, $w_2$,$w_3$ and $w(y)=y(1-y)^2$. The light (heavy) 
dotted line corresponds to the ALEPH fit (our fit) for the weight $w^{(00)}$,
the light (heavy) dashed line to the ALEPH fit (our fit) for the
weight $w_2$, the light (heavy) dot-dashed line to the ALEPH fit (our fit)
for the weight $w_3$, and the light (heavy) double-dot-dashed line to
the ALEPH fit (our fit) for the weight $y(1-y)^2$.
The right boundary corresponds to the kinematic endpoint, 
$s_0=m_\tau^2\simeq 3.16\ {\rm GeV}^2$.}
\rotatebox{270}{\mbox{
\begin{minipage}[thb]{11.7cm}
\begin{picture}(11.6,14.1)
\epsfig{figure=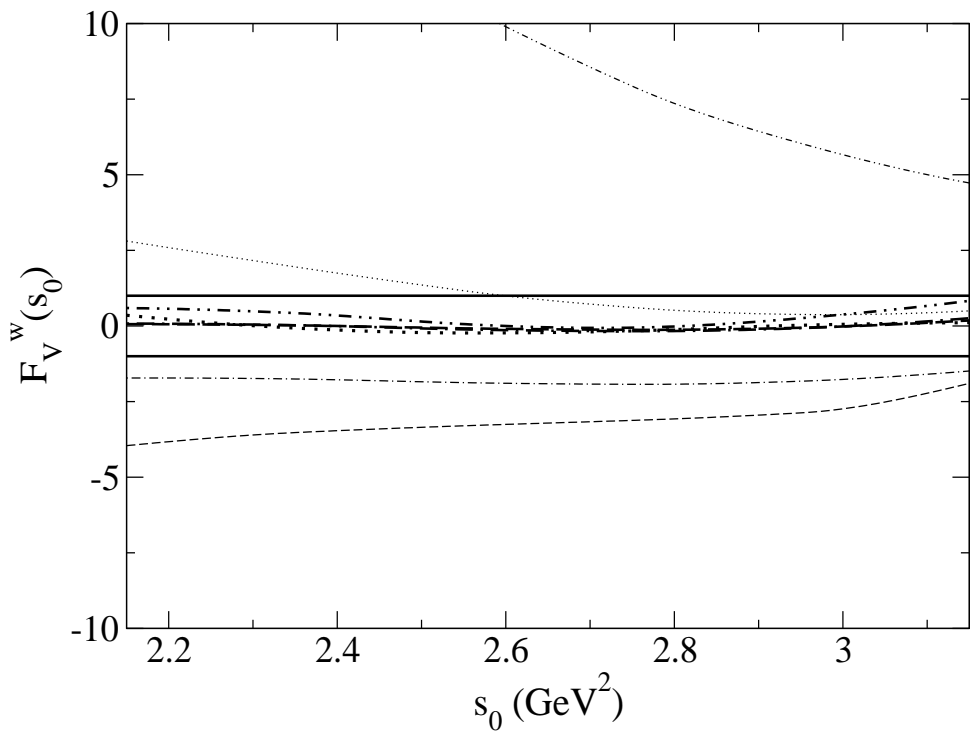,height=14.5cm,width=11.5cm}
\end{picture}
\end{minipage}}}
\label{fitqualcomparisonfig}\end{figure}

Further evidence for the reliability of our fits for $\alpha_s$
and the $C_D$ is provided by the fact that, unlike the fit qualities
associated with the ALEPH fit parameter sets, those associated
with our fits remain between $-1$ and $1$ for all three channels,
all five $w_N$, and all $s_0$ in our fit window. This is illustrated for the
V channel in Fig.~\ref{fitqualcomparisonfig}, which shows the $F^w_V(s_0)$ 
corresponding to our fits (denoted by the heavy lines) for the four weights 
discussed above ($w^{(00)}$, $w_2$, $w_3$ and $w(y)=y(1-y)^2$) whose OPE
integrals do not depend on any of the $C_{D>8}$. Also shown, for comparison 
are the corresponding ALEPH fit results (denoted by the light lines) 
for this same set of weights and same channel, 
shown previously in Fig.~\ref{fitqualfig}. The comparison makes
evident the major improvement represented by our fit results.
One might argue that the much improved fit quality in the $w_2$ and $w_3$ cases
is a result of the fact that our parameters were obtained by fitting
to the corresponding spectral integrals. The excellent quality of
the fit to the $w^{(00)}$- and $y(1-y)^2$-weighted spectral
integrals, however, is a strong test of the implicit assumption
that the form assumed on the OPE side of our FESRs in fact correctly
incorporates all relevant OPE contributions, an assumption already
shown to fail for the more restrictive forms assumed in the
earlier combined spectral weight analyses. We remind the reader that the
suppression of the $D=0$ contribution for the $w(y)=y(1-y)^2$
case makes the agreement in that case an even more significant
test of the reliability of the $C_6$ and $C_8$ values obtained
using the $w_2$ and $w_3$ FESRs. 

The situation in the V+A channel, which is the source
of our central $\alpha_s$ determination, is similar
to that found in the V channel. Specifically, we find 
\begin{itemize}
\item $\vert F_{V+A}^{w_N}(s_0)\vert$ for our optimized 
fits even smaller than those found in the V channel (and hence also
uniformly $<1$ over the whole of the fit window employed); 
\item $\vert F_{V+A}^{w_{(km)}}(s_0)\vert$ corresponding to the 
2005 ALEPH optimized fit typically $>>1$ away from $s_0=m_\tau^2$;
\item $\vert F_{V+A}^{w}(s_0)\vert$ results produced by applying
the optimized 2005 ALEPH values for the $D\leq 8$ OPE fit parameters
to the degree $\leq 3$ weights $w_2(y)$, $w_3(y)$ and $y(1-y)^2$ not 
employed in the ALEPH fit significantly $>1$, even for $s_0=m_\tau^2$; and
\item in contrast, $\vert F_{V+A}^{w}(s_0)\vert$ results produced by applying
our optimized $D\leq 8$ OPE fit parameters to the degree $\leq 3$ weights 
$w_{(00)}$ and $y(1-y)^2$ not employed in our fits 
uniformly $<1$ through the region of the fit window employed.
\end{itemize}
In view of the similarity to the V channel results, we do not
provide explicit analogues of Figures~\ref{fitqualfig} and
~\ref{fitqualcomparisonfig} for the V+A channel.


\subsubsection{$D=0$ Convergence}
The next point for discussion is the pattern of convergence of the results
for $\alpha_s$ with increasing truncation order. This is relevant to the
question of the extent to which our estimate for the $D=0$ truncation 
uncertainty is a conservative one. In Table~\ref{alphaslooporder},
we display the results for $\alpha_s(m_\tau^2)$ obtained from full fits to
the ALEPH-based V+A $w_N$ FESRs as a function of the truncation 
order, $M$, in $\alpha_s$, employed for the $D=0$ series. The extremely 
good consistency (to within $\pm 0.0001$ across the set of $w_N$ employed) 
allows us to quote a single common value for each truncation order. 
The behavior of the extracted values of $\alpha_s(m_\tau^2)$ with
increasing $M$ appears reasonable and, we would claim, 
supports the interpretation of our
truncation uncertainty estimate of $\pm 0.010$ on $\alpha_s(m_\tau^2)$ 
as a sensibly conservative one. For comparison, the scheme for
estimating the truncation uncertainty employed in 
Ref.~\cite{davieretal08} produces the less conservative 
assessment ${}^{+0.0062}_{-0.0074}$.

\begin{table}
\caption{\label{alphaslooporder}The fitted values for $\alpha_s(m_\tau^2)$ 
obtained from the ALEPH-based $w_N$-weighted V+A analyses 
as a function of the $D=0$ truncation order, $M$, where $M$ here
specifies that the last term kept in the $D=0$ series for the Adler function 
is that proportional to $d_M[\alpha_s(Q^2)]^M$. Our central analyses
above correspond to $M=5$.}
\vskip .1in
\begin{tabular}{|c|cccc|}
\hline
$M$&2&3&4&5\\
\hline
$\alpha_s(m_\tau^2)$&$0.375$&$0.338$&$0.326$&$0.320$\\
\hline
\end{tabular}
\end{table}

\subsubsection{Comparisons to other determinations of the $D>4$ parameters,
$C_D$}
We turn now to the issue of the extracted values of the
$D>4$ condensate combinations, making comparisons to other 
determinations of these same combinations appearing in the
literature. The analysis above is, of course, designed specifically
to reduce $D>4$ OPE contributions and, as such, is far from 
optimal for the determination of the $C_D$. As a result, the precision
in our determinations of most of the $C_D$ is not high. In 
Table~\ref{cdcomparison} we compare our results (with the experimental 
and theoretical errors now combined in quadrature) with those of ALEPH, 
OPAL and two other recent condensate studies~\cite{dscD07,aas08}, focussing
on the quantities $C_{6,8}$ obtained in those earlier studies. In the ALEPH and
OPAL cases, the errors shown are the nominal ones quoted in the original 
publications, and do not include the sizeable additional uncertainty 
associated with the neglect of $D>8$ contributions
discussed already above.
In the case of Ref.~\cite{dscD07},
which employs fits using the weights $w(y)=1-y^N$ (which have a zero
of order $1$ at $y=1$), we quote only the values considered reliable by
the authors themselves, and of these, only the ones corresponding to 
$\Lambda =350$ MeV, since it is this value which lies closest
to that ($346$ MeV) associated with our central fit result above.
In the case of Ref.~\cite{aas08} we quote only the A channel
$C_6$ result, since this was the only one to display
demonstrable stability, within errors, in going from the 
$2$-parameter fit (including contributions up to $D=6$) 
to the $3$-parameter fit (including contributions up to 
$D=8$)~\cite{aas08comment}.  

\begin{table}
\caption{\label{cdcomparison}Comparison of our results for
$C_6$ and $C_8$ with those of Refs.~\cite{davieretal08} (ALEPH),
\cite{opalud99} (OPAL), \cite{dscD07} (DS) and \cite{aas08} (AAS).
$C_6$ is given in units of $10^{-3}$ GeV$^6$ and $C_8$ in 
units of $10^{-3}$ GeV$^8$. The errors
quoted are as described in the text.} 
\vskip .1in
\begin{tabular}{|l|cc|cc|cc|}
\hline
Reference&$C_6^V$&$C_8^V$&$C_6^A$&$C_8^A$&$C_6^{V+A}$&$C_8^{V+A}$\\
\hline
ALEPH&\ \ $-3.6(3)$\ \ &\ \ $5.0(3)$\ \ &\ \ $4.6(3)$\ \ 
&\ \ \ $-6.0(3)\ $\ \ &\ \ $1.0(5)$\ \ &\ \ $-1.0(5)$\ \ \\
OPAL&\ \ $-3.4(5)$\ \ &\ \ $5.0(8)$\ \ &\ \ $2.6(5)$\ \ 
&\ \ $-2.6(1.3)$\ \ &\ \ $-0.3(1.5)$\ \ &\ \ $1.3(4.2)$\ \ \\
DS&\ \ $-8.9(3.0)$&---&\ \ $-4.3(3.0)$\ \ 
&---&---&---\\
AAS&---&---&\ \ $-2.4(2.0)$\ \ 
&---&---&---\\
\hline
Our fit&\ \ $-5.9(2.0)$\ \ &\ \ $6.0(7.0)$\ \ &\ \ $-2.3(2.0)$\ \ 
&\ \ $18.1(7.6)$\ \ &\ \ $-8.4(3.8)$\ \ &\ \ $25.1(13.2)$\ \ \\
\hline
\end{tabular}
\end{table}

We note that, for the V channel, where the ALEPH fit quality was
better, our $C_8$ values actually agree well with those of ALEPH and OPAL,
while our $C_6$ central values are somewhat larger, but of the
same general size. For the A channel, where the ALEPH fit quality 
was poorer, we have, instead, significant disagreement for $C_6$, not just
in magnitude, but also in the sign of the central value. 
The significant differences for the A channel are also seen in the 
V+A channel, as one would expect. Since our values lead to extremely good OPE
representations for the $w^{(00)}$, $w_2$, $w_3$ and
$w(y)=y(1-y)^2$ spectral integrals in all three channels, while the
ALEPH and OPAL fits do not, it is no surprise that significant
differences between our fits and theirs should be found. We note that the
disagreement in sign for $C^A_6$ confirms the result found
in Refs.~\cite{dscD07,aas08}. As pointed out in those references, the
fit results imply a significant breakdown of the vacuum saturation 
approximation (VSA) for the four-quark $D=6$ condensates, since VSA values for
the V and A channel are in the ratio $-7:11$. 
While it is true that, given the size of the errors, the sign of $C_6^A$ is
not firmly established by either our fits or those of 
Refs.~\cite{dscD07,aas08}, nonetheless the relative magnitudes 
of the V and A results are far from satisfying the VSA relation.
To improve on the accuracy of the determinations of the $C_D$,
and investigate such issues further, would require working with a
different set of weight functions, chosen in such a way as to suppress
$D=0$ and emphasize higher $D$ contributions.

\subsection{\label{subsecfinalresults}Final results}
In order to avoid the additional uncertainties associated with the 
separation of the observed V+A spectral distribution 
into its V and A components, we base our final results 
for $\alpha_s$ on the V+A $w_N$ FESR analyses. 
As seen above, the agreement of the ALEPH- and OPAL-based V+A results
is excellent. The individual ALEPH V and A fits are, 
in addition, in extremely good agreement with the corresponding V+A results, 
though, of course, with larger experimental errors. The agreement of the 
ALEPH V, A and V+A central values is considerably 
closer than that obtained from the spectral weight analysis of 
Ref.~\cite{davieretal08}. It should be stressed that the agreement in
the present case is obtained using the value of $\langle aG^2\rangle_{RGI}$ 
determined independently in Ref.~\cite{newgcond4}, in sharp contrast 
to the A and V+A fits of Ref.~\cite{davieretal08}, which require
incompatible, and unambiguously negative, values.

Averaging the V+A results, using the non-normalization
component of the experimental errors, we obtain
\begin{equation}
\alpha_s(m_\tau^2)\, =\, 0.3209(46)(118)
\label{finalalphastau}\end{equation}
where the first error is experimental (now including the normalization
uncertainty) and the second theoretical. The experimental error is
identical to that obtained in the spectral weight analysis of
Ref.~\cite{davieretal08}, while our theoretical error is larger
as a result of the more conservative treatment of the
$D=0$ truncation uncertainty. The theoretical error of the earlier analyses, 
of course, does not include the additional contribution
identified above, associated with the neglect of $D>8$ OPE contributions.

The $n_f=5$ result, $\alpha_s(M_Z^2)$, is obtained from the
$n_f=3$ result given in Eq.~(\ref{finalalphastau}) using the standard
self-consistent combination of $4$-loop running with $3$-loop matching at 
the flavor thresholds~\cite{cks97}. As shown in Ref.~\cite{bck08}, taking 
$m_c(m_c)=1.286(13)$ GeV and $m_b(m_b)=4.164(25)$ GeV~\cite{kss07}, 
the matching thresholds to be $rm_{c,b}(m_{c,b})$ with
$r$ varying between $0.7$ and $3$, and incorporating
uncertainties associated with the truncated running and matching,
produces a combined evolution uncertainty of $0.0003$ on $\alpha_s(M_Z^2)$. 
Our final result is then
\begin{equation}
\alpha_s(M_Z^2)\, =\, 0.1187(3)(6)(15)
\label{finalalphasmz}\end{equation}
where the first uncertainty is due to evolution, the second is experimental
and the third theoretical. The difference between this value and
that obtained in the earlier spectral weight analysis,
$0.1212(11)$, serves to quantify the impact of the 
$D>8$ contributions neglected in the previous analysis.

The result, Eq.~(\ref{finalalphasmz}), is in good agreement with 
a number of recent independent experimental determinations, specifically,
\begin{itemize}
\item the 2008 updates of the global fit to electroweak observables at the 
$Z$ scale, quoted in Refs.~\cite{bck08,davieretal08}, which yield
$\alpha_s(M_Z^2)=0.1190(26)$ and $0.1191(27)_{exp}(1)_{th}$, respectively;
\item the combined NLO fit to the inclusive jet cross-sections measured
by H1 and ZEUS~\cite{alphasherajetcrosssections}, which yields 
$\alpha_s(M_Z^2)=0.1198(19)_{exp}(26)_{th}$;
\item the NLO fit to high-$Q^2$ $1$-, $2$- and $3$-jet cross-sections 
measured by H1 (presented at DIS 2008 and the 2008 HERA-LHC 
workshop~\cite{alphash1jetshighqsq08}) which yields
$\alpha_s(M_Z^2)=0.1182(8)_{exp}\left( ^{+41}_{-31}\right)_{scales}(18)_{pdf}$;
\item the NNLO fit to event shape observables in $e^+ e^-\rightarrow hadrons$
at LEP~\cite{alphaseventshapes08},
which yields $\alpha_s(M_Z^2)=0.1240(33)$; 
\item the SCET analyis, including resummation of next-to-next-to-next-to
leading logarithms, of ALEPH and OPAL thrust distributions
in $e^+ e^-\rightarrow hadrons$~\cite{bs08}, which yields 
$\alpha_s(M_Z^2)=0.1172(13)_{exp}(17)_{th}$; and
\item the fit to $e^+ e^-\rightarrow hadrons$ cross-sections between
$2$ GeV and $10.6$ GeV CM energy~\cite{kst07}, which yields 
$\alpha_s(M_Z^2)=0.119\left( ^{+9}_{-11}\right)$.
\end{itemize}
The agreement with the recent updated analysis of 
$\Gamma [\Upsilon (1s)\rightarrow \gamma X]/
\Gamma [\Upsilon (1s)\rightarrow X]$~\cite{brambillaetalupsilon07}, 
which replaces the older analysis usually cited in the PDG QCD 
review section, and yields $\alpha_s(M_Z^2)=0.119\left( ^{+6}_{-5}\right)$,
is also good.
Note that the $\tau$ decay extraction is considerably more precise than 
any of the other experimental determinations. In addition,
the $\tau$ decay and lattice results, whose discrepancy was noted
at the outset, are now seen to be compatible within errors. This
compatibility is, in fact, further improved by the increase in
$\alpha_s(M_Z)$ found in two recent studies~\cite{hpqcd08,mlms08} 
which revisit the earlier lattice determination, incorporating lattice 
data at a wider range of scales than that employed in 
Ref.~\cite{latticealphas}.

\subsection{Some Comments on the Recent Beneke-Jamin Study
and Its Relation to the Present Work}
After the completion of the work described in this paper,
a new exploration of the extraction of $\alpha_s$ from hadronic
$\tau$ decay data was posted~\cite{jb08}. This study employs
a 5-parameter model for the Borel transform of the $D=0$ component
of the Adler function, one whose structure incorporates the 
form of the known leading UV renormalon and two leading IR renormalon 
singularities. The parameters of the model are fixed using the 
known coefficients, $d^{(0)}_,\cdots ,d^{(0)}_4$, of the $D=0$ Adler 
function series expansion, together with the estimated value $d^{(0)}_5=283$. 
The study makes the working assumption that the true all-orders result 
will be well approximated by the Borel 
sum of the corresponding model Adler function series. The results 
generated using the model are then
argued to favor the use of FOPT over CIPT for the
$D=0$ OPE contribution. It is not clear to us whether 
extended ansatze for the Borel transform, involving
additional parameters, would lead to the same or 
different conclusions. We do comment, however, that the 
results for $\alpha_s(m_\tau^2)$ obtained from our FOPT fits, though yielding 
representations of the spectral integral data which are of nearly as 
good quality as those produced by the corresponding CIPT fits, 
are significantly less consistent than those obtained using the CIPT prescription,
the results for the V+A channel ranging from $0.320$ for $w_2$ to $0.312$ for
$w_6$. Whether one views this as an empirical argument in favor 
of softening the conclusions of Ref.~\cite{jb08} or not, the
arguments of that reference clearly support taking a conservative
approach to assessing the $D=0$ truncation uncertainty. 

\begin{figure}
\unitlength1cm
\caption{The fit qualities $F^{w}_{V+A}(s_0)$ corresponding to
the ALEPH data, the OPE parameters of Ref.~\cite{jb08}, and the
FOPT evaluation of the $D=0$ OPE contributions, for the $w^{(00)}$, 
$w_2$, $w_3$ and $w(y)=y(1-y)^2$ FESRs. The dotted, dashed, dot-dashed and 
double-dot-dashed lines correspond to $w^{(00)}$, $w_2$, $w_3$ and $y(1-y)^2$,
respectively. The right boundary corresponds
to the kinematic endpoint, $s_0=m_\tau^2\simeq 3.16\ {\rm GeV}^2$.}
\rotatebox{270}{\mbox{
\begin{minipage}[thb]{11.7cm}
\begin{picture}(11.6,14.1)
\epsfig{figure=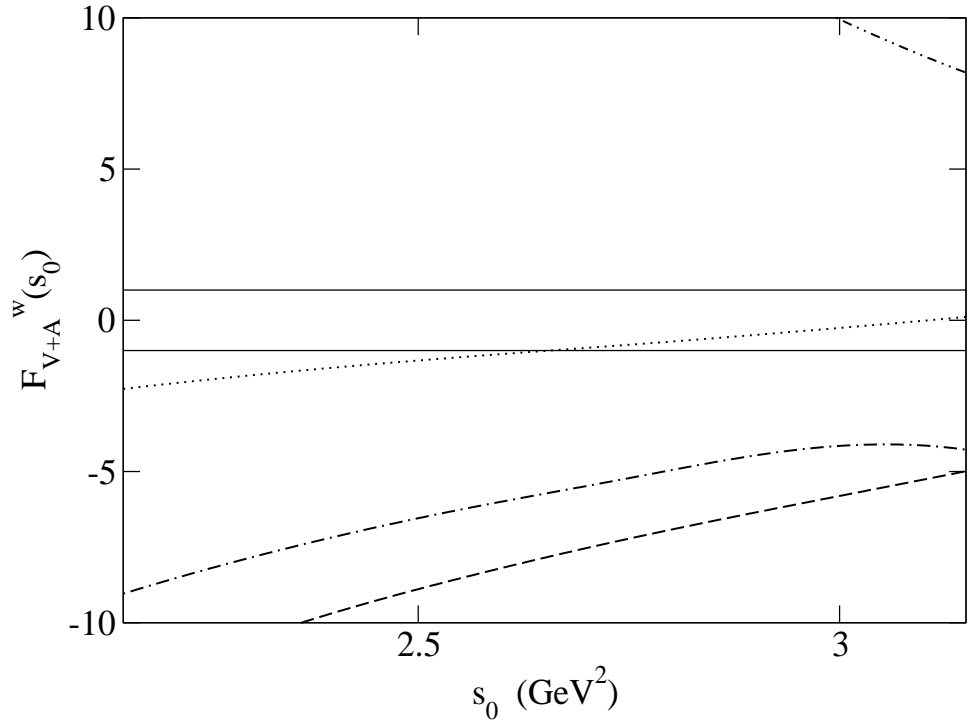,height=14.5cm,width=11.5cm}
\end{picture}
\end{minipage}}}
\label{jb08vpa_comparison_jun08}\end{figure}

For readers inclined to adopt the FOPT determination as the central one
(in spite of the reduced consistency of its output), we comment
that the $\alpha_s(m_\tau^2)$ obtained from
the $w_2$ through $w_6$ V+A fits correspond to values of
$\alpha_s(M_Z^2)$ lying between $0.1186$ and $0.1176$.
The CIPT result, as it turns out, not only displays better
consistency, but is also in better agreement with the 
results reported in Refs.~\cite{hpqcd08,mlms08}, which
update the original lattice analysis of Ref.~\cite{latticealphas}.

Regarding the values for $\alpha_s(m_\tau^2)$ and $\alpha_s(M_Z^2)$ quoted 
in Ref.~\cite{jb08}, the reader should bear in mind that these result from 
a $w^{(00)}$-weighted V+A FESR analysis restricted to the single value 
$s_0=m_\tau^2$. With only a single $s_0$, it is not possible to fit
$C_6^{V+A}$ and $C_8^{V+A}$, and central values (and errors)
must therefore be assumed for these quantities. The authors
of Ref.~\cite{jb08} take the central value for $C^{V+A}_6$ 
to be given by twice the VSA result and that for $C^{V+A}_8$ to be $0$.
Our fifth order FOPT fits in fact return significantly different values.

It is possible to test the consistency of the assumed values for 
$C_6^{V+A}$ and $C_8^{V+A}$ with the resulting extracted value of 
$\alpha_s(m_\tau^2)$, as above, by studying the $s_0$-dependence 
of the match between the OPE and spectral integral sides of the 
$w_2$, $w_3$, $w^{(00)}$ and $w(y)=y(1-y)^2$ FESRs, whose OPE sides 
do not depend on any of the $C_{D>8}$. The reader, here, should bear in mind
that, in Ref.~\cite{jb08}, slightly different values of $d_5^{(0)}$ and 
$\langle a G^2\rangle_{RGI}$ were employed than those used above.
Using the $d_5^{(0)}$, $\langle a G^2\rangle_{RGI}$, $C_6^{V+A}$ and 
$C_8^{V+A}$ values of Ref.~\cite{jb08}, together with the resulting
$O(\bar{a}^5)$-truncated FOPT fit value for $\alpha_s(m_\tau^2)$,
we find the fit qualities, $F_{V+A}^w(s_0)$, 
displayed in Fig.~\ref{jb08vpa_comparison_jun08}.
$F_{V+A}^{w^{(00)}}(s_0)$ is, of course, small
near $s_0=m_\tau^2$ since the value of $\alpha_s(m_\tau^2)$ employed
in the calculations was fixed using the $s_0=m_\tau^2$ version of 
the $w^{(00)}$ FESR. The deterioration in the fit quality for $w^{(00)}$
as $s_0$ is decreased, as well as the very poor fit qualities
for the other three weights, clearly demonstrates that
the values assumed for $C_6^{V+A}$ and $C_8^{V+A}$ are problematic. 
The value obtained for $\alpha_s(m_\tau^2)$ using these values
as input should thus also be treated with caution. We have already noted 
the results of our own FOPT fits above. Since the $\alpha_s(m_\tau^2)$
values obtained from the $w_2$ and $w_3$ FESRs do not show the
same degree of consistency as was observed in the CIPT-based fit,
it would be necessary to perform a combined fit, using a number of
the degree $\leq 3$ weights, to improve further on the FOPT determination.

\subsection{Final summary and comments}
To summarize, we have performed a number of related FESR analyses
designed specifically to reduce the impact of poorly known
$D>4$ OPE contributions on the extraction of $\alpha_s$ using
hadronic $\tau$ decay data. Our results show a high degree of 
consistency and satisfy constraints not satisfied by other $\tau$ decay
determinations. Our final result is 
\begin{equation}
\alpha_s(M_Z^2)\, =\, 0.1187\pm 0.0016
\label{finalalphasmzcombined}\end{equation}
where the evolution, experimental and theoretical errors have now been 
combined in quadrature. The result is in excellent agreement with (and more 
precise than) alternate independent high-scale experimental determinations.
It is, however, significantly lower than the values obtained 
in the earlier ALEPH and OPAL hadronic $\tau$ decay analyses. 
We have provided clear evidence that the source of
this discrepancy lies in the contamination of these earlier 
combined spectral weight analyses by
neglected, but non-negligible, $D>8$ OPE contributions.

A technical point worth emphasizing from the discussion above is
the importance of working with a range of $s_0$ rather than just the single
value $s_0=m_\tau^2$, and the utility, in this context, 
of using weights defined in terms of the dimensionless variable $y=s/s_0$.
For such weights, the $s_0$-dependence of the 
resulting weighted spectral integrals allows one to 
straightforwardly test any assumptions made
about the values of $D>4$ OPE coefficients, or, better yet, 
to attempt actual fits to obtain these values using data. 
Such $s_0$-dependence studies seem to us unavoidable if one wishes to 
demonstrate that $D>4$ OPE contributions have indeed been brought under 
control at the level ($\sim 0.5\%$ of the full spectral integrals)
required for a $\sim 1\%$ precision determination of $\alpha_s(M_Z^2)$.
Fortunately, as we have shown, such control
is not difficult to achieve, and we have displayed a number of
weights which are useful for this purpose. The weights, $w_N(y)$,
which isolate individual integrated $D=2N+2$ contributions,
are related to the kinematic weight, $w^{(00)}(y)$,
by slowly varying multiplicative factors~\cite{multfactorfootnote},
and hence produce errors on the spectral integrals that are
comparable to, or better than, those for $w^{(00)}$. 

We stress that theoretical errors now dominate the
uncertainty in the hadronic $\tau$ decay determination of $\alpha_s(M_Z^2)$, 
the $D=0$ OPE truncation error being the largest among these. 
Further reduction in experimental errors, and in particular, 
improvements in the V/A separation, are likely to be possible 
using data from the B factories, and such improvements
would be useful for further testing the consistency of the V, A and
V+A determinations. Given the current situation, however, 
reduced experimental errors would have little impact on
the total error on $\alpha_s(M_Z^2)$.

\begin{acknowledgments}
KM would like to acknowledge the hospitality of the CSSM, University of
Adelaide, the ongoing support of the Natural Sciences and Engineering 
Council of Canada, and useful interchanges with Matthias Jamin. 
TY wishes to acknowledge helpful discussions with
R. Koniuk and C. Wolfe while this work was in progress.
\end{acknowledgments}


\vfill\eject


\begin{thebibliography}{99}
\bibitem{sewfootnote}In what follows, we neglect 
the small additive electroweak correction, usually denoted 
$\delta^\prime_{EW}$ in the literature, which is known explicitly only
for the $(00)$ spectral weight case, and for $s_0=m_\tau^2$~\cite{bl90}.
In that case it is at the sub-$0.1\%$ level relative to the leading $D=0$ 
OPE contribution, a level which would produce a shift in $\alpha_s$ much
smaller than the main sources of error in our determination.
\bibitem{alephkkbarpifootnote}Note that while in 
Ref.~\cite{davieretal08} ALEPH has provided the $K\bar{K}\pi$ contribution 
to the isovector V spectral distribution, the covariance information
needed to make the corresponding improvement to the errors on the
V and A spectral integrals is not publicly available. We will thus
quote experimental errors based on the uncorrected, but publicly
available, covariances for the V, A inclusive sums.
\bibitem{dimlessfootnote}As will be discussed in more detail later
in the text, the reason for working with a range of $s_0$,
and with weights which are functions of the dimensionless variable $y$,
is that certain internal consistency checks on the treatment of higher $D$
OPE contributions then become possible. It turns out that these checks
are crucial to achieving a reliable high precision determination of
$\alpha_s(M_Z^2)$.
\bibitem{higherdlogsfootnote}In the interests of brevity, we will suppress 
the phrase ``up to $\alpha_s$-suppressed logarithmic corrections'' 
in what follows. Such a reminder 
should, however, be understood as being implicitly present in any statement 
concerning integrated $D>4$ OPE contributions.
\bibitem{higherdlogsdetailfootnote}Note that higher order corrections to 
the leading order $D>4$ OPE contributions to the correlator begin
with terms of the form $C^{(1)}_D \alpha_s(Q^2)/Q^D$. These
can be re-expressed by expanding $\alpha_s(Q^2)$ as a power series, e.g., in 
$\alpha_s^0\equiv \alpha_s(m_\tau^2)$: 
$\alpha_s(Q^2)=\alpha_s^0+\sum_{k=2}c_k(t)\left[\alpha_s^0\right]^k$,
where the $c_k(t)$ are polynomials in $t=log(Q^2/m_\tau^2)$ whose
coefficients are determined by those of the $\beta$ function. The
first order term in the resulting expansion of the $\alpha_s$-suppressed 
correlator contribution, $C^{(1)}_D\alpha_s^0/Q^D$, can thus be absorbed 
into the leading order term $C_D/Q^D$ through a redefinition of the 
coefficient $C_D$. This part of the higher order $D>4$ contribution is 
thus automatically included whenever $C_D$ is fitted to data. Higher order 
integrated $D>4$ ``logarithmic corrections'' incorporated only in an
average sense as part of the fitting procedure thus begin with a 
term proportional to $\left[ \alpha_s^0\right]^2\, log(s_0/m_\tau^2)$.
\bibitem{bck08footnote}It should be noted that the analysis of
Ref.~\cite{bck08}, which employs the $w^{(00)}$ spectral weight, 
uses the 2005 ALEPH fits for the coefficients $\delta^{(6)}$
and $\delta^{(8)}$ as input. Problems with these fit values thus
also impact the reliability of that analysis.
\bibitem{wrongdlogsfootnote}Note that, for reasons entirely analogous to
those discussed in detail above~\cite{higherdlogsdetailfootnote}, 
integrated OPE contributions to the $w_N$-weighted FESR having 
dimension $D\not= 2N+2$ begin at $O\left( [\alpha_s(s_0)]^2\right)$.
There is also a numerical suppression as a result of the structure
of the relevant weighted logarithmic integrals, combined with
the alternation in signs of the polynomial coefficients associated
with the pinching condition $w(y=1)=0$. This strong suppression
of $D\not= 2N+2$ contributions is of course also shared by
the weights $w(y)=1-y^N$, a feature of these weights
emphasized in Ref.~\cite{dscD07}.
\bibitem{aas08comment}Since no fit incorporating $d=4,6,8$ and $10$
contributions is reported for the V, A or V+A channels, it is not
possible to check the $D=8$ results of Ref.~\cite{aas08} for
stability. For this reason, combined with the observed instability
of the $C_6$ fits for the V and V+A channels, we do not quote the
$D=8$ results.
\bibitem{multfactorfootnote}If we write $w_N(y)=w^{(00)}(y)c_N(y)$,
$c_N(0)=1$ and $c_N(1)=N/6$. $c_2$, $c_3$ and $c_4$ decrease monotically 
on $[0,1]$, while $c_5$ reaches a minimum, $\sim 0.774$ at 
$y\sim 0.551$ and $c_6$ a minimum, $\sim 0.803$ at $y\sim 0.432$.
\end{thebibliography}
\end{document}